\documentclass[%
superscriptaddress,
notitlepage,
showpacs,preprintnumbers,
nofootinbib,
 amsmath,amssymb,
 aps,
]{revtex4-1}
\usepackage[dvipdfmx]{graphicx}
\usepackage{dcolumn}
\usepackage{bm}
\usepackage{hyperref}
\usepackage{comment}
\usepackage{color}

\newcommand{\Mare}[1]{\textcolor{magenta}{#1}}
\newcommand{\KA}[1]{\textcolor{cyan}{\bf#1}}

\begin{document}

\preprint{IPMU 19-0101}

\title{Super-sample tidal modes on the celestial sphere}

\author{Kazuyuki Akitsu}
\affiliation{%
 Kavli Institute for the Physics and Mathematics of the Universe (WPI),\\
 The University of Tokyo Institutes for Advanced Study (UTIAS),\\
 The University of Tokyo, Chiba 277-8583, Japan
}%
\affiliation{%
 Department of Physics, Graduate School of Science,\\
 The University of Tokyo, 7-3-1 Hongo, Bunkyo-ku, Tokyo 113-0033, Japan
}%

\author{Naonori S. Sugiyama}
\affiliation{
National Astronomical Observatory of Japan, Mitaka, Tokyo 181-8588, Japan
}

\author{Maresuke Shiraishi}
\affiliation{%
 Department of General Education, National Institute of Technology, Kagawa College, 355
 Chokushi-cho, Takamatsu, Kagawa 761-8058, Japan
}%


\date{\today}

\begin{abstract}

The super-sample tidal effect carries information on long-wavelength fluctuations that we cannot measure directly.
It arises from the mode-coupling between short-wavelength and long-wavelength perturbations beyond a finite region of a galaxy survey
and violates statistical isotropy of observed galaxy power spectra.
In this paper, we propose the use of bipolar spherical harmonic (BipoSH) decomposition formalism to characterize statistically anisotropic power spectra.
Using the BipoSH formalism, we perform a comprehensive study of the effect of the super-sample tides on measurements of other cosmological distortions such as
the redshift-space distortion (RSD) and Alcock-Paczy\'{n}ski (AP) effects by means of the Fisher information matrix formalism.
We find that the BipoSH formalism can break parameter degeneracies among the super-sample tidal, RSD and AP effects,
indicating that the super-sample tides have little impact on the measurements of the RSD and AP effects.
We also show that the super-sample tides are detectable with an accuracy better than the $\Lambda$CDM prediction without 
impairing the accuracy of measurements of other anisotropies assuming a SPHEREx-like galaxy survey.

\end{abstract}
\maketitle

\section{\label{sec:intro}Introduction}

The large-scale structure (LSS) of the Universe offers a powerful tool for measuring the cosmic expansion history of the Universe.
As the LSS keeps information on the initial state of the Universe, its measurements 
can also be used to test the fundamental statistical properties of cosmic fluctuations predicted by inflation~\cite{Starobinsky:1980,Sato:1980,Guth:1980}.
Given the great success of the SDSS III BOSS project~\cite{Alam_etal:2016}, 
various next-generation galaxy redshift surveys such as Prime Focus Spectrograph (PFS)~\cite{PFS_wp}, Large Synoptic Survey Telescope (LSST)~\cite{LSST_wp}, Dark Energy Spectroscopic Instrument (DESI)~\cite{DESI_wp} and Spectro-Photometer for the History of the Universe, Epoch of Reionization, and Ices Explorer (SPHEREx)~\cite{Spherex_wp}
are ongoing and planned.
For interpreting upcoming unprecedentedly high-quality data correctly, 
it is of crucial importance to accurately model various non-linear corrections imprinted in the observed galaxy clustering:
non-linear gravitational instabilities~\cite{PT_review}, non-linear galaxy biases~\cite{Bias_review} and non-linear redshift-space distortions~\cite{Scoccimarro:2004}.

The non-linear growth of the LSS produces the mode-coupling of different scales. 
The mode-coupling naturally predicts that long-wavelength fluctuations beyond a given survey region may affect the observed galaxy clustering within a finite survey region, which is known as the super-sample or super-survey effect~\cite{Takada_Hu:2013}.
We cannot directly measure these long-wavelength fluctuations (called as the super-sample or super-survey modes) in a finite volume survey.
However, through the non-linear mode coupling between different Fourier modes, the super-sample modes change both the amplitude and the comoving scale of the short-wavelength fluctuations, which is known as the growth and dilation effect, respectively~\cite{Hamilton_etal:2005,Sherwin_Zaldarriaga:2012,Li_etal:2014a}.

The Effects of the super-sample modes on the real-space power spectrum have been extensively studied in Refs.~\cite{Hamilton_etal:2005,dePutter_etal:2011,Takada_Hu:2013,Li_etal:2014a,Li_etal:2014b,Akitsu_etal:2016,Barreira_etal:2017,Schmidt_etal:2018}.
The uncertainty of the amplitude of the super-sample modes forces us to add the new term to the power spectrum covariance, dubbed as the super-sample covariance~\cite{Takada_Hu:2013,Li_etal:2014a,Barreira_etal:2017}.
Physical effects of the super-sample modes originate from the second derivatives of large-scale gravitational potential,
which can be decomposed into the trace (mean overdensity) part and the traceless (large-scale tidal field) one~\cite{Akitsu_etal:2016}.
Thus, there are super-sample tidal components that are expected to be of the same order of magnitude as those of isotropic ones,
whereas many of previous studies have focused mainly on the isotropic super-sample mode
because the impact of super-sample tidal components on the real-space power spectrum vanishes after spherical averages.

An observed clustering pattern of galaxies is anisotropically distorted by the peculiar velocity of galaxies along the line-of-sight (LOS), known
as the redshift-space distortions (RSD)~\cite{Kaiser:1987}.
An additional anisotropic signal arises due to the (mis)transformation from observed quantities into comoving distances.
Converting an observed redshift and polar position on the sphere, $(z,\theta)$,
into a radial and tangential comoving distance, $(x_\parallel, x_\perp)$, requires the use of the fiducial cosmological parameters;
$x_\parallel= z/H(z) $ and $x_\perp = (1+z)D_A(z)\theta$.
Then, if the assumed cosmological parameters differ from the underlying true cosmological parameters,
an apparent anisotropic distortion along the LOS is induced, 
which is known as the Alcock-Paczy\'{n}ski (AP) effect~\cite{Alcock_Paczynski:1979}.
Besides these well-known effects, 
the super-sample tidal modes also generate a new anisotropic signature in the redshift-space galaxy power spectrum
~\cite{Akitsu_etal:2016,Akitsu_Takada:2017,Li_etal:2017,Chiang_Slosar:2018}. 
These similar effects tend to give rise to degeneracies in cosmological parameter estimation \cite{Akitsu_Takada:2017,Chiang_Slosar:2018}.

To break the parameter degeneracy, in the past few years people have started to use extra degrees of freedom of the observed galaxy power spectrum, i.e., \textit{violation of statistical isotropy}.
The super-sample tidal perturbation generates a preferred direction in a given local survey region 
and breaks statistical isotropy~\cite{Schmidt_etal:2013,Akitsu_etal:2016,Zhu_etal:2015,Barreira_Schmidt:2017}.
The anisotropic distortion induced by the RSD and AP effects, where the statistical isotropy still holds, is characterized by an angle between the wave vector $\textbf{k}$ and the LOS unit vector $\hat{n}$
and thus can entirely be decomposed using the Legendre polynomials ${\cal L}_{\ell}(\hat{k}\cdot\hat{n})$. 
In order to extract information on the breaking of statistical isotropy due to the super-sample tidal modes, \citet{Chiang_Slosar:2018} proposed an expansion scheme of the three-dimensional power spectrum with spherical harmonics functions.
The authors decomposed the $\textbf{k}$-dependence according to $P(\textbf{k},\hat{n}) = \sum_{\ell m} P_{\ell m}(k) Y_{\ell m}(\hat{k})$ after the LOS direction $\hat{n}$ is defined as a $z$-axis.
Note that the $m=0$ mode corresponds to the coefficient in the normal Legendre expansion scheme since $Y_{\ell 0} \propto {\cal L}_\ell$. They found that the signals due to the RSD effect are confined to $m = 0$, while the tidal perturbation creates non-vanishing $m \neq 0$ modes.
The authors further performed a Fisher matrix computation and showed that their decomposition formalism can break the degeneracy between the RSD effect and the super-sample tidal one
except for its LOS component.

In this paper, we examine the distinguishability between the super-sample tidal effect and the other two ones (the RSD and AP effects) by employing a more general decomposition based on bipolar spherical harmonics (BipoSH) 
$\{Y_{\ell}(\hat{k})\otimes Y_{\ell'}(\hat{n}) \}_{LM}$ \cite{book:Varshalovich:1988}. 
This was recently applied to probing primordial statistical anisotropy induced by some sort of vector inflation models~\cite{Shiraishi_etal:2016,Sugiyama_etal:2017,Bartolo:2017sbu}.
  \footnote{The BipoSH decomposition was initially introduced for dealing with the wide-angle effect in the power spectrum \cite{Szalay:1997cc, Szapudi:2004gh,Papai:2008bd,Bertacca:2012tp,Raccanelli:2013dza}. For an application to the galaxy bispectrum analysis, See Ref.~\cite{Sugiyama:2018yzo}.}

Through this decomposition, statistically anisotropic signals are confined to the $L \neq 0$ BipoSH coefficients. We here follow the methodology developed in Ref.~\cite{Shiraishi_etal:2016}, and, differently from Ref.~\cite{Chiang_Slosar:2018}, we do not fix $\hat{n}$ to any specific direction.
This treatment is reasonable for actual data analysis
because it is impossible to determine a global LOS direction $\hat{n}$ in observed galaxy samples.
In Ref.~\cite{Sugiyama_etal:2017}, the BipoSH formalism was already applied to observed galaxy samples in order to constrain statistically anisotropic signals. 
There, the effects of observational systematics, e.g., artificial asymmetries due to specific survey geometry, were also decomposed and hence properly subtracted.
The same data analysis pipeline will also be applicable to the measurements of the super-sample tidal modes. 

Via the BipoSH decomposition of the redshift galaxy power spectrum, it is confirmed that only the super-sample tidal effect induces non-vanishing $L = 2$ coefficients. 
Moreover, using these BipoSH coefficients, we perform a Fisher matrix computation and forecast the detectability of relevant cosmological parameters.
In the Fisher matrix, we include the contributions of not only the super-sample tidal and RSD effects but also the AP one, which was unconsidered in Ref.~\cite{Chiang_Slosar:2018}, and find that the super-sample tidal effect has little impact on estimates of both the RSD and AP effects.

This paper is laid out as follows.
In Section \ref{sec:preliminariy}, 
we review the effect of the super-sample modes on the galaxy power spectrum in redshift space and the formulation of the bipolar spherical harmonics (BipoSH) expansion.
In Section \ref{sec:fisher}, we show the results of Fisher forecasts for the parameters which characterize the super-sample tides and other distortion parameters.
We discuss some applications and conclude in Section \ref{sec:conclusion}.
In Appendix \ref{app:relation}, we give the relations between the BipoSH expansion used in this paper and Legendre expansion, which usually used in the RSD analysis, and between the BipoSH expansion and the single spherical harmonic expansion used in Ref.~\cite{Chiang_Slosar:2018}.
In Appendix \ref{app:sss_biposh}, we provide the details of calculations of the BipoSH multipoles.
Mathematical identities used for computations are summarized in Appendix \ref{app:Identitis}.

\section{\label{sec:preliminariy}Preliminaries}

The primary goal of this paper is to investigate how the super-sample effect
contaminates the RSD and AP effect on the observed galaxy power spectrum by using the full three-dimensional information.
In Section~\ref{subsec:effect_SS} we review the galaxy power spectrum including the super-sample effect at leading order.~\cite{Akitsu_Takada:2017}.
In Section~\ref{subsec:BipoSH}, we explain the (BipoSH) decomposition formalism of the three-dimensional power spectrum. 
Then, we explicitly show the BipoSH coefficients for the redshift-space galaxy power spectrum with the full super-sample modes.

\subsection{\label{subsec:effect_SS}Effects of the super-sample modes on the observed power spectrum of galaxies in redshift space}
Since we observe galaxies within a finite survey region, 
the observed density fluctuation, $\delta_{\rm obs}$, is represented as the convolution of the survey window function, $W$, and the true density fluctuation, $\delta$:
\begin{align}
    \delta_{\rm obs}({\bf k}) \equiv \int \frac{d^3 {\bf q}}{(2\pi)^3} W({\bf k}-{\bf q}) \delta({\bf q}),
\end{align}
where $W(\bf q)$ is the Fourier counterpart of the survey window function $W({\bf x})$.
Note here that the observed density fluctuation has non-zero value even at ${\bf k}=\bf{0}$ because of the survey window function. 
Throughout this paper, we refer to the $\bf{k} = \bf{0}$ mode of $\delta_{\rm obs}({\bf k})$ as the "large-scale overdensity", given by~\cite{Takada_Hu:2013}
\begin{align}
    \delta_{\rm b} \equiv \int \frac{d^3 {\bf q}}{(2\pi)^3} W(-{\bf q}) \delta({\bf q}).
\end{align}
Furthermore, it is convenient to define the large-scale tidal components as~\cite{Akitsu_etal:2016}
\begin{align}
    \tau_{ij} \equiv \int \frac{d^3 {\bf q}}{(2\pi)^3} W(-{\bf q}) \left(\hat{q}_i \hat{q}_j -\frac{1}{3}\delta^K_{ij} \right) \delta({\bf q}),
\end{align}
where $\hat{q}_i\equiv q_i/q$ with $q=|{\bf q}|$, and $\delta^K_{ij}$ is the Kronecker delta.

The expectation values of $\delta_{\rm b}$ and $\tau_{ij}$ are zero: $\langle \delta_{\rm b} \rangle =\langle \tau_{ij} \rangle =0$.
On the other hand, their variances can be computed as follows
\begin{align}
    &\sigma^2_{\rm b} = \langle \delta^2_{\rm b} \rangle 
    =\frac{1}{V^2}\int \frac{d^3 {\bf q}}{(2\pi)^3}P_{\rm lin}(q)|W({\bf q})|^2,
    \label{eq:variance_delta}
    \\
    &\langle \tau_{ij}\tau_{\ell m} \rangle 
    =\frac{1}{V^2}\int\frac{d^3 {\bf q}}{(2\pi)^3} \left(\hat{q}_i\hat{q}_j -\frac{1}{3}\delta^K_{ij} \right)
                                                        \left(\hat{q}_{\ell}\hat{q}_m -\frac{1}{3}\delta^K_{\ell m} \right) P_{\rm lin}(q)|W({\bf q})|^2
    \nonumber\\
    &\hspace{1.22cm}= \left(-\frac{2}{45}\delta^K_{ij}\delta^K_{\ell m}+\frac{3}{45}\delta^K_{i\ell}\delta^K_{j m}+\frac{3}{45}\delta^K_{im}\delta^K_{j \ell} \right)\sigma_{\rm b}^2,
    \label{eq:variance_tau}
\end{align}
where $V$ is a survey volume, and $P_{\rm lin}(q)$ is the linear matter power spectrum.
In the second equality in Eq.~(\ref{eq:variance_tau}), we assumed an isotropic window function, $W({\bf q})=W(q)$, for simplicity.

As shown above, the window function picks up the longer-wavelength fluctuations than a typical scale of survey volume.
Throughout this paper, we assume that a given survey volume is so large that the super-sample modes grow linearly, and therefore, $|\delta_{\rm b}|,|\tau_{ij}|\ll 1$.
Although $\delta_{\rm b}$ and $\tau_{ij}$ are related through $\tau_{ij}=\partial_i\partial_j \partial^{-2} \delta_{\rm b}$ in real space,
the values of $\tau_{ij}$ cannot be inferred from $\delta_{\rm b}$ due to the non-local nature of a tidal field (suggested by the appearance of inverse Laplacian $\partial^{-2}$, see Ref.~\cite{Dai_etal:2015,Ip_Schmidt:2016} for details).
Then we have six independent degrees of freedom for the super-sample modes: one isotropic component $\delta_{\rm b}$ and five anisotropic components $\tau_{ij}$.
Because the super-sample modes depend on the position and shape of specific surveys and we cannot predict these values for each survey,
we need to vary these six components as free parameters in cosmological analyses, as will be studied in detail in Section~\ref{sec:fisher}.

\subsubsection{Power spectrum responses to the super-sample modes}

These long-wavelength perturbations affect the small-scale clustering due to the nonlinear mode-coupling by gravity.
In the presence of the super-sample modes, the galaxy power spectrum is modulated as~\cite{Takada_Hu:2013,Akitsu_etal:2016}
\begin{align}
    P_g({\bf k},\hat{n}; \delta_{\rm b}, \tau_{ij})
    = P_g({\bf k},\hat{n}; \delta_{\rm b}=0, \tau_{ij}=0) + \frac{\partial P_g(k,\mu)}{\partial \delta_{\rm b}}\delta_{\rm b} 
                    + \frac{\partial P_g({\bf k},\hat{n})}{\partial \tau_{ij}}\tau_{ij} 
                    +\mathcal{O}(\delta_{\rm b}^2 ,\tau^2_{ij})
\end{align}
where $\partial P_g(k,\mu)/\partial \delta_{\rm b}$ and 
$\partial P_g({\bf k},\hat{n})/\partial \tau_{ij}$ are the \textit{responses} of the galaxy power spectrum to the $\delta_{\rm b}$ and $\tau_{ij}$
respectively,
which represent the scale\KA{-}dependent modulation to the observed power spectrum caused by the super-sample modes 
and $\mu$ is the cosine between the wave vector ${\bf k}$ and the line-of-sight (LOS) $\hat{n}$.
Notice that the response to the isotropic super-sample mode $\delta_{\rm b}$ can depend on only $k$ and $\mu$ 
because it preserves the rotational symmetry around the observer.
This mode-coupling between long- and short-wavelength modes is then characterized by the squeezed limit of the bispectrum~\cite{Akitsu_Takada:2017,Barreira_Schmidt:2017,Li_etal:2017}.
In particular, the responses are related to the squeezed bispectrum
\cite{Akitsu_Takada:2017}
\begin{align}
\lim_{q\to 0}B_{ggm}(\textbf{k},-\textbf{k}-\textbf{q},\textbf{q})
=\left[\frac{\partial P_g(\textbf{k})}{\partial \delta_\textrm{b}} +\left(\hat{q}_i\hat{q}_j -\frac{1}{3}\delta^K_{ij} \right)\frac{\partial P_g(\textbf{k})}{\partial \tau_{ij}} \right],
\label{eq:squeezed_B}
\end{align}
with $\langle \delta_{g}({\bf k_1})  \delta_{g}({\bf k_2}) \delta_{m}({\bf q}) \rangle\equiv 
B_{ggm}({\bf k_1},{\bf k_2},{\bf q})(2\pi)^3\delta_D^{(3)}({\bf k_1}+{\bf k_2}+{\bf q})$, $\delta_g({\bf k})$ being
the overdensity field of galaxies and $\delta_m({\bf k})$ being
the overdensity field of matters.

As can be seen in Eq.~(\ref{eq:squeezed_B}), one can obtain the explicit form of the response functions from the squeezed bispectrum.
Using the standard perturbation theory~\cite{PT_review}, the tree-level squeezed bispectrum in redshift space is expressed as \cite{Akitsu_Takada:2017,Barreira_Schmidt:2017}
\begin{align}
    B_{ggm}({\bf k}, -{\bf k}-{\bf q}, {\bf q}) = 2Z_1({\bf k+q})Z_2({\bf k+q},-{\bf q})P^L(|{\bf k+q}|)P^L(q) 
                                                +2Z_1({\bf k})Z_2({\bf k},{\bf q})P^L(k)P^L(q),   
    \label{eq:squeezed_B_pert}
\end{align}
where the kernel functions~\cite{PT_review}
\begin{align}
    Z_1({\bf k}) =& b_1 + f\mu^2_{k},
    \\
    Z_2({\bf k_1},{\bf k_2}) =&b_1 F_2({\bf k_1},{\bf k_2}) + \frac{b_2}{2} + \frac{b_{s^2}}{2}\left( (\hat{k}_1 \cdot \hat{k}_2) -\frac{1}{3} \right)
        +f\mu^2_{K}G_2({\bf k_1},{\bf k_2}) 
        \nonumber\\
        &\hspace{1cm}- \frac{f\mu_{K}K}{2}\left[\frac{\mu_{k_1}}{k_1}(b_1+f\mu_{k_1}^2)+\frac{\mu_{k_2}}{k_2}(b_1+f\mu_{k_2}^2) \right]
\end{align}
are the mode-coupling kernels in redshift space with $b_1$, $b_2$, and $b_{s^2}$ being bias parameters up to the second order, $f$ being the growth rate, ${\bf K}\equiv {\bf k}_1+ {\bf k}_2$,
$\mu_{k} = \hat{k}\cdot \hat{n}$ and 
\begin{align}
    F_2({\bf k_1},{\bf k_2}) =&\frac{5}{7}+\frac{1}{2}\left( \frac{1}{k_1^2}+\frac{1}{k_2^2} \right)({\bf k}_1 \cdot {\bf k}_2) + \frac{2}{7}(\hat{k}_1\cdot\hat{k}_2)^2
    \\
    G_2({\bf k_1},{\bf k_2}) =&\frac{3}{7}+\frac{1}{2}\left( \frac{1}{k_1^2}+\frac{1}{k_2^2} \right)({\bf k}_1 \cdot {\bf k}_2) + \frac{4}{7}(\hat{k}_1\cdot\hat{k}_2)^2.
\end{align}
By comparing terms in Eqs.~(\ref{eq:squeezed_B}) and (\ref{eq:squeezed_B_pert}), the response of the redshift-space galaxy power spectrum to the large-scale overdensity $\delta_{\rm b}$ is read off as~\cite{Akitsu_Takada:2017,Chiang_Slosar:2018}
\begin{align}
\frac{\partial P_g(k ,\mu)}{\partial \delta_{\rm b}}
=&
\left[ \frac{47}{21}b_1+2b_2-\frac{1}{3}b_1\frac{d \ln{P_{\rm lin}(k)}}{d \ln{k}}\right]b_1 P^L(k)
+\left[\frac{1}{3}b_1^2 +\mu^2 \left(\frac{26}{7}b_1 +2b_1^2+2b_2\right) -\frac{\mu^2}{3}b_1(2+b_1)\frac{d \ln{P_{\rm lin}(k)}}{d \ln{k}} \right] f P^L(k)\nonumber \\
&+\left[\frac{1}{21}(31+70b_1)-\frac{1}{3}(1+2b_1)\frac{d \ln{P_{\rm lin}(k)}}{d \ln{k}} \right]f^2 \mu^4 P^L(k)
+\left[\frac{1}{3}(4\mu^2-1)-\frac{1}{3}\mu^2\frac{d \ln{P_{\rm lin}(k)}}{d \ln{k}} \right]f^3 \mu^4 P^L(k),
\end{align}
and to the large-scale tides $\tau_{ij}$ as~\cite{Akitsu_Takada:2017,Chiang_Slosar:2018}
\begin{align}
\frac{\partial P_g({\bf k},\hat{n})}{\partial \tau_{ij}}
=&\left[\frac{8}{7}b_1+2b_{s^2}-b_1\frac{d \ln{P_{\rm lin}(k)}}{d \ln{k}} \right]\hat{k}_i\hat{k}_j b_1 P^L(k)
\nonumber\\
&+\left[
b_1^2 \hat{n}_i\hat{n}_j 
+ \frac{24}{7}b_1\mu^2 \hat{k}_i\hat{k}_j 
+2b_{s^2}\mu^2
-b_1\mu \left(2\mu \hat{k}_i\hat{k}_j+b_1 h_{ij} \right)\frac{d \ln{P_{\rm lin}(k)}}{d \ln{k}}
\right] f P^L(k)
\nonumber\\
&+\left[
\frac{16}{7}\mu \hat{k}_i\hat{k}_j 
+4b_1 h_{ij}
- \left(\mu \hat{k}_i\hat{k}_j+2b_1 h_{ij} \right)\frac{d \ln{P_{\rm lin}(k)}}{d \ln{k}}
\right] \mu^3 f^2 P^L(k)
\nonumber\\
&+\left[
\left(
4\mu h_{ij} 
-\hat{n}_i\hat{n}_j 
\right)
-\mu
h_{ij}
\frac{d \ln{P_{\rm lin}(k)}}{d \ln{k}}
\right] \mu^4 f^3 P^L(k),
\end{align}
where $h_{ij} \equiv k_{(i}n_{j)} = \frac{1}{2}(k_i n_j + n_i k_j)$.
In the limit $f \to 0$,
the above equations reduce to the real-space results~\cite{Schmidt_etal:2013,Akitsu_etal:2016}.

The expressions above formulate the physical effects of the super-sample modes on small-scale fluctuations.
There are two types of the super-sample effects.
First, the super-sample modes enhance or suppress the growth of the short-modes depending on the sign of the super-sample modes and directions of small-scale fluctuations:
speeding up the growth in the denser region and slowing down in the less dense region.
This growth effect corresponds to the terms with no derivatives.
Second, the super-sample modes cause a dilation of the comoving scale since the local expansion history is altered by the super-sample modes.
The mean density mode $\delta_{\rm b}$ generates an isotropic shift for all scales.
On the other hand, the tidal modes $\tau_{ij}$ cause an ellipsoidal expansion in a local region and this leads to an anisotropic shift which depends on the directions of both the LOS and the wave vector of the short-modes.
This dilation effect is described by the derivative terms.

In particular, this dilation leaves a characteristic imprint on the baryon acoustic oscillation (BAO) feature in the power spectrum.
Specifically, the isotropic super-sample mode shifts the observed BAO scale in an isotropic way, whereas 
the anisotropic super-sample modes shift in an anisotropic way.
Neglecting the growth terms, we can rewrite the redshift-space galaxy power spectrum with responses as
\begin{align}
    P_g({\bf k}, \hat{n};\delta_{\rm b}, \tau_{ij})
    =& (b_1 + f\mu^2)^2 P_{\rm lin}(k) +  \left(\mathcal{D}(\mu)\delta_{\rm b} + \mathcal{D}_{ij}(\hat{k},\hat{n})\tau_{ij}\right)\frac{\partial P_{\rm lin}(k)}{\partial \ln{k}}
    \nonumber \\
    \simeq& (b_1+f\mu^2)^2 P_{\rm lin}\left(k/\alpha(\hat{k},\hat{n})\right),
\end{align}
where $\mathcal{D}(\mu)$ and $\mathcal{D}_{ij}(\hat{k},\hat{n})$ are the coefficients of the dilation term ($k$-derivative term) for $\delta_{\rm b}$ and $\tau_{ij}$ respectively and 
 \begin{align}
    \alpha(\hat{k},\hat{n})  
    =& \left[1  + \frac{\mathcal{D}(\mu)}{(b_1+f\mu^2)^2}\delta_{\rm b} + \frac{\mathcal{D}_{ij}(\hat{k},\hat{n})}{(b_1+f\mu^2)^2}\tau_{ij} + \mathcal{O}(\delta^2_{\rm b},\tau^2_{ij}) \right]^{-1}
    \nonumber\\
    \simeq& 1 - \frac{1}{3}(1+f\mu^2)\delta_{\rm b} - (\hat{k}_i\hat{k}_j + f\mu h_{ij})\tau_{ij}
\end{align}
parameterizes the direction-dependent shift in the BAO peak.
When there is no BAO peak shift, $\alpha =1$ holds.
From this expression, one can easily see that the large-scale tides generate three-dimensionally anisotropic distortions in the BAO peak position,
while the large-scale mean density causes only two-dimensional distortion.

\subsubsection{Modulation of the mean galaxy overdensity}
In the previous subsection, we implicitly assume that the overdensity field of galaxies is defined to the global (background) mean number density of galaxies.
In a spectroscopic survey of galaxies, however, we measure the overdensity field defined to the ``local" mean number density in the survey region.
Because the super-sample modes behave like the background in the local survey area, these also make a difference between the ``local" mean number density $\bar{n}^{\rm global}_g$ and the ``global" mean number density $\bar{n}^{\rm global}_g$ such that $\bar{n}^{\rm global}_g = \bar{n}^{\rm local}_g (1+\Delta_g) $ with $\Delta_g$ being the mean galaxy overdensity in the specific survey due to the super-sample modes.
In a galaxy redshift survey, therefore, the observed number density fluctuation of galaxies $\delta_g^{\rm local} ({\bf k})$ which is defined through $n_g({\bf k})=\bar{n}^{\rm local}_g[1+\delta^{\rm local}_g({\bf k})]$ is related to that defined to the ``global" mean density through $n_g({\bf k})=\bar{n}_g^{\rm global}[1+\delta_g^{\rm global} ({\bf k})]$ as follows~\cite{dePutter_etal:2011},
\begin{align}
    \delta^{\rm local}_g ({\bf k})
    = \frac{\delta_g^{\rm global}({\bf k})}{1+\Delta_g}
    \simeq (1 -\Delta_g ) \delta_g^{\rm global} ({\bf k}).
\end{align}
In redshift space $\Delta_g$ is related to the super-survey modes,
\begin{align}
    \Delta_g = \left[ b_1 + f (\hat{q}\cdot\hat{n})^2 \right] \delta_{\rm b} = \left( b_1 + \frac{1}{3}f \right) \delta_{\rm b} + f\tau_{ij} \hat{n}^i \hat{n}^j
\end{align}
at lowest order \cite{Li_etal:2017,Chiang_Slosar:2018}.
This means that the observed power spectrum of galaxies is modulated as 
\begin{align}
    P^{\rm local}_g({\bf k},\hat{n}) 
    \simeq & (1 - 2 \Delta_g) P_g^{\rm global} ({\bf k},\hat{n})
    =  \left[1 -  2\left( b_1 + \frac{1}{3}f \right) \delta_{\rm b} - 2f\tau_{ij} \hat{n}^i \hat{n}^j\right] P_g^{\rm global} ({\bf k},\hat{n}),
\end{align}
where
\begin{align}
    P_g^{\rm global} ({\bf k},\hat{n})
    = (b_1 + f\mu^2 )^2 P_{\rm lin}(k)
    + \frac{\partial P_g (k,\mu)}{\partial \delta_{\rm b}}\delta_{\rm b}
    + \frac{\partial P_g ({\bf k},\hat{n})}{\partial \tau_{ij}}\tau_{ij}.
\end{align}
After all, at leading order of the super-sample modes, the observed power spectrum of galaxies with the effects of the super-sample modes is expressed as 
\begin{align}
    P^{\rm local}_g({\bf k},\hat{n}) = (b_1 + f\mu^2)^2 P_{\rm lin}(k)
    &+\left[ -2\left(b_1+\frac{1}{3}f\right)(b_1+f\mu^2)P_{\rm lin}(k)+ \frac{\partial P_g ( k,\mu)}{\partial \delta_{\rm b}}\right]\delta_{\rm b}
    \nonumber\\
    &+\left[-2f(b_1+f\mu^2)^2P_{\rm lin}(k)\hat{n}^i\hat{n}^j + \frac{\partial P_g ({\bf k},\hat{n})}{\partial \tau_{ij}}\right]\tau_{ij}.
    \label{eq:3D_power}
\end{align}
We use this power spectrum in the Fisher analysis.

\subsection{\label{subsec:BipoSH}The bipolar spherical harmonic expansion}
The power spectrum which depends upon two directions, $\hat{k}$ and $\hat{n}$, can be expressed by using the following coordinates,
\begin{align}
    {\bf k} =& k (\sin{\theta_k}\cos{\phi_k},~ \sin{\theta_k}\sin{\phi_k},~ \cos{\theta_k})
    \\
    \hat{n} =&  (\sin{\theta_n}\cos{\phi_n},~ \sin{\theta_n}\sin{\phi_n},~ \cos{\theta_n}).
\end{align}
In general, to get the multiple moments that have no angular dependence requires a four-multiple integration, which is the case for the BipoSH expansion as we will see in the next section.
Note that the reason why we usually need only one-dimensional integral for the power spectrum multipoles in redshift space is that 
the usual RSD anisotropy still preserves the three-dimensionally rotational symmetry around the observer.
In that case, the four-multiple integration reduces to one-dimensional integral thanks to the rotational symmetry.
We here emphasis that, differently from Ref.~\cite{Chiang_Slosar:2018}, throughout this paper, the LOS direction is not set to the global one although the local plane-parallel approximation is adopted.

\subsubsection{Formalism of the bipolar spherical harmonic expansion}
To capture the violation of the three-dimensional rotational symmetry,
the three-dimensional power spectrum $P^s({\bf k},\hat{n}; \delta_{\rm b}, \tau_{ij})$ in redshift space
should be expanded using the bipolar spherical harmonic (BipoSH) basis $X^{LM}_{\ell\ell'}$~\cite{Shiraishi_etal:2016},
\begin{align}
P^s({\bf k},\hat{n};\delta_{\rm b},\tau_{ij})
&=\sum_{LM \ell\ell '}
\pi^{LM}_{\ell\ell '}(k;\delta_{\rm b},\tau_{ij})
X^{LM}_{\ell\ell '}(\hat{k},\hat{n}),
\end{align}
where the BipoSH basis is defined as~\cite{book:Varshalovich:1988}
\begin{align}
X^{LM}_{\ell\ell '}(\hat{k},\hat{n})
&=\{Y_{\ell}(\hat{k})\otimes Y_{\ell'}(\hat{n}) \}_{LM}
=\sum_{mm'} \mathcal{C}^{LM}_{\ell m \ell 'm'}
Y_{\ell m }(\hat{k})Y_{\ell ' m '}(\hat{n})\\
&=\sum_{mm'}
(-1)^{\ell-\ell ' +M}\sqrt{2L+1}
\begin{pmatrix}
\ell & \ell '  & L\\
m & m' & -M
\end{pmatrix}
Y_{\ell m }(\hat{k})Y_{\ell ' m '}(\hat{n}),
\end{align}
with the $\mathcal{C}^{LM}_{\ell m \ell 'm'}$ being the Clebsch-Gordan coefficients and  
$\left(\begin{smallmatrix}
\ell_1 & \ell_2  & \ell_3\\
m_1 & m_2 & m_3
\end{smallmatrix}\right)$
being the Wigner 3$j$ symbol. The inverting translation is given by
\begin{align}
\pi^{LM}_{\ell\ell '}(k)
=\int d^2\hat{k}
\int d^2\hat{n}~
P({\bf k},\hat{n})
[X^{LM}_{\ell\ell '}(\hat{k},\hat{n})]^*,
\label{eq:BipoSH_integral}
\end{align}
owing to the orthogonal property of the $X^{LM}_{\ell\ell '}$ basis:
\begin{align}
\int d^2\hat{k}
\int d^2\hat{n}~
X^{L_1M_1}_{\ell_1\ell_1 '}(\hat{k},\hat{n})
[X^{L_2M_2}_{\ell_2\ell_2 '}(\hat{k},\hat{n})]^*
=\delta^K_{L_1 L_2}\delta^K_{M_1 M_2}
\delta^K_{\ell_1 \ell_2}\delta^K_{\ell_1' \ell_2'}.
\end{align}
To relate the coefficients of the bipolar spherical harmonic expansion with those of usual Legendre expansion, let us introduce the \textit{reduced} coefficients defined in Refs.~\cite{Shiraishi_etal:2016,Sugiyama_etal:2017}
\begin{align}
P_{\ell \ell'}^{LM}(k)
\equiv \pi^{LM}_{\ell \ell'}(k) (-1)^L \frac{\sqrt{(2L+1)(2\ell+1)(2\ell'+1)}}{4\pi}H_{\ell \ell' L},
\end{align}
where
$H_{\ell_1 \ell_2 \ell_3} \equiv 
\left(\begin{smallmatrix}
\ell_1 & \ell_2  & \ell_3\\
0 & 0 & 0
\end{smallmatrix}\right)$ 
and $H_{\ell\ell'L}=0$ when $\ell+\ell'+L$ is odd. 
In our case, $\pi^{LM}_{\ell \ell'}$ vanishes for $\ell + \ell' + L = \rm odd$ and hence $P^{LM}_{\ell\ell'}$ is sufficient to capture all information of three-dimensional power spectrum.
The explicit relationships between the \textit{reduced} coefficients $P_{\ell\ell'}^{LM}(k)$ and the usual Legendre coefficients $P_{\ell}(k)$ and between $P_{\ell\ell'}^{LM}(k)$ and the coefficients obtained via the expansion scheme of Ref.~\cite{Chiang_Slosar:2018}
is present in Appendix \ref{app:relation}.

The advantages of employing the BipoSH expansion lies in the following two aspects:
(1) The BipoSH can extract the full three-dimensional anisotropic power spectrum
and (2) we need not set the LOS to the $z$-axis.
The reason why the point (1) is important is that the RSD and AP distortions generate only two-dimensional asymmetry; i.e. anisotropic signature appears only about the LOS direction,
characterized by the radial components of the wave vector, $\hat{k}_{\parallel} = \hat{n}$,
whereas the super-sample tidal effect sources the full three-dimensional asymmetry; i.e. anisotropic imprint appears not only about the LOS direction but also in the transverse plane,
characterized by both $\hat{k}_{\parallel}$ and $\hat{k}_{\perp} =\widehat{{\bf k} - {\bf k}_{\parallel}}$.

To put the point (2) another way, the use of the BipoSH expansion requires the multiple LOS directions as implied by the integration of the LOS direction 
(see Eq.~(\ref{eq:BipoSH_integral})).
In other words, the BipoSH expansion can be applied in an all-sky or wide-area survey.
In fact, different LOS directions are essential to break the degeneracies between the super-sample tidal effect and other anisotropic effects.
The point is that the RSD and AP distortion respect the rotational symmetry around the observer, 
whereas the super-sample tidal modes violate the rotational invariance.

\begin{figure}[t]
	\centering
    \includegraphics[width=8cm]{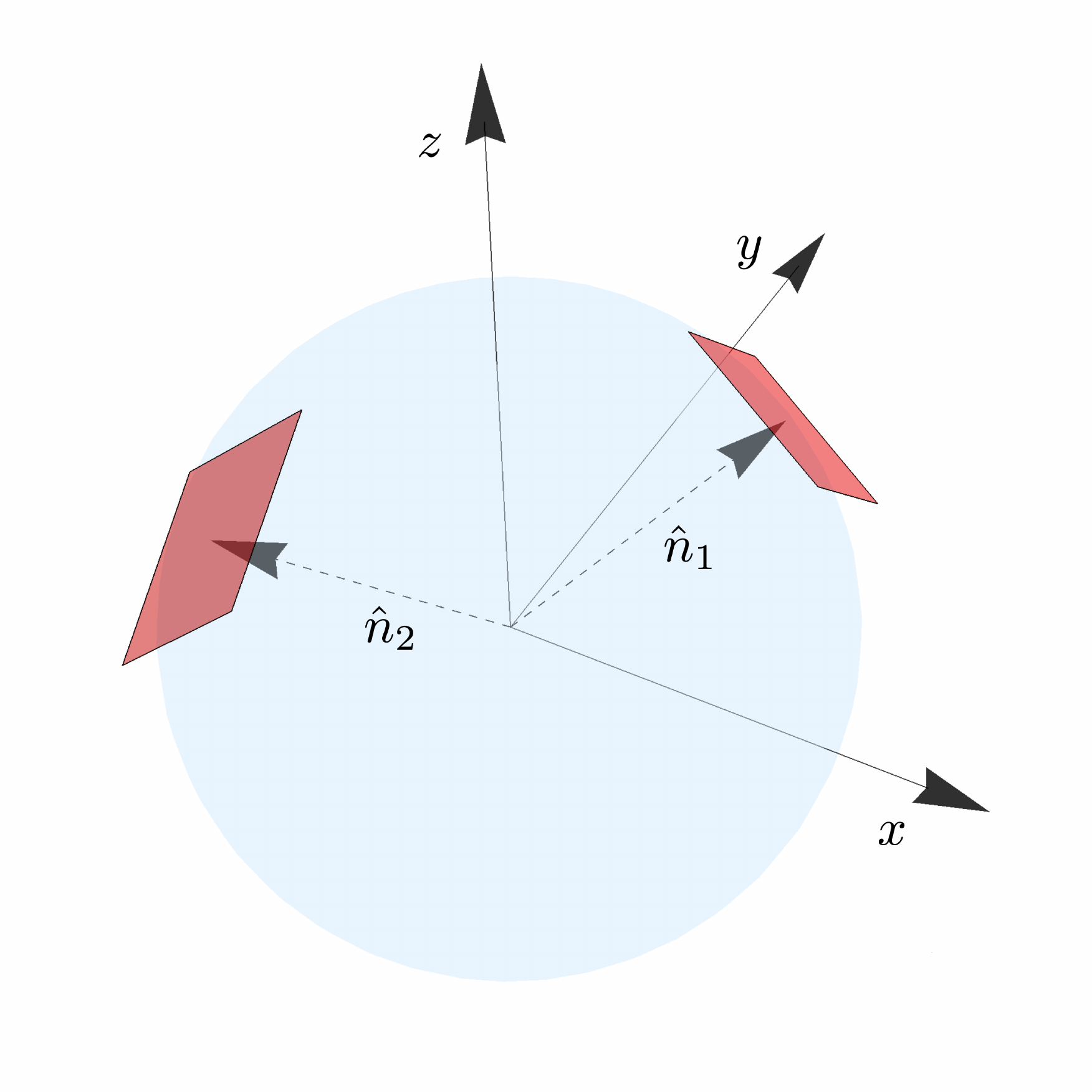}
	\caption{A schematic picture for an all-sky galaxy survey. The observer is at the origin.
	$n_{1}= (1,1,1)/\sqrt{3}$ and $n_2= (-1,-1,1)/\sqrt{3}$ depict different line-of-sight (LOS) direction. 
	In the local plane parallel approximation, the pairs of galaxies are measured in each red plane, which is the tangential plane to each LOS direction.}
	\label{fig:sphere}
\end{figure}

To elucidate this point further, let us consider a simplified situation where we have two different LOS; $\hat{n}_1 = (1,1,1)/\sqrt{3}$ and $\hat{n}_2 = (-1,-1,1)/\sqrt{3}$ (see Fig.~\ref{fig:sphere}).
Under the local plane-parallel approximation,
we observe the galaxy pairs on the tangential plane for each LOS (the red planes in Fig.~\ref{fig:sphere}).
For both $\hat{n}_1$ and $\hat{n}_2$ direction, we should measure the same power spectrum in the absence of $\tau_{ij}$
because in that case anisotropy appeared in the power spectrum depends only $\mu=(\hat{k}\cdot\hat{n})$, 
which is rotationally invariant around the observer.
On the other hand,
the terms with super-sample tidal modes violate this rotational symmetry.
For example, let us consider the term of $\tau_{ij}\hat{n}_i\hat{n}_j$.
This anisotropy appears with the form of $2(\tau_{12}+\tau_{13}+\tau_{23})/3$ for $\hat{n}_1$ 
and $2(\tau_{12}-\tau_{13}-\tau_{23})/3$ for $\hat{n}_2$,
which means that we would observe the power spectrum with the different radial distortion for each different LOS.
The separation of $P^{00}_{\ell\ell}$ and $P^{2M}_{\ell\ell'}$ originates from this fact as we will see in the next subsection.

In summary, if one performs angler average in each tangential plane by which the two-dimensional power spectrum is obtained and set $\hat{n}=\hat{z}$,
only the information on the radial distortion is left, and the fact that $\tau_{ij} \hat{n}_1^i \hat{n}_1^j \neq \tau_{ij} \hat{n}_2^i \hat{n}_2^j $ 
obliges us to introduce different parameters which describe the super-sample tides for various LOS.
On the other hand, since the BipoSH expansion captures the full three-dimensional power spectrum including the information of the distortion on the tangential planes and taking into account the non-parallel LOS,
it is expected to alleviate the degeneracies between the super-sample tidal effect and other anisotropic effects.
Notice that here $\tau_{33}$ is no longer the LOS component of $\tau_{ij}$; i.e. $\tau_{33}\neq\tau_{ij}\hat{n}^i\hat{n}^j$, instead $zz$ component in the coordinate the observer chooses; $\tau_{33}=\tau_{ij}\hat{z}^i\hat{z}^j$.

\subsubsection{BipoSH coefficients of the response functions}
By making use of the BipoSH 
formalism, we can decompose Eq.~(\ref{eq:3D_power}) into the following $reduced$ coefficients,
\begin{align}
    P^{00}_{\ell\ell'}(k) &= \delta^K_{\ell\ell'} \left[P_{\ell}(k) + D_{\ell}(k) P_m(k) \delta_{\rm b}\right]
    \\
    P^{20}_{\ell\ell'}(k) &= T_{\ell\ell'}(k) P_m(k) \tau_{33}
    \label{Eq:BipoSH_M0}
    \\
    P^{2\pm 1}_{\ell\ell'}(k) &= T_{\ell\ell'}(k) P_m(k) \sqrt{\frac{2}{3}} (\mp\tau_{13}+i\tau_{23})
    \label{Eq:BipoSH_M1}
    \\
    P^{2\pm 2}_{\ell\ell'}(k) &= T_{\ell\ell'}(k) P_m(k) \frac{1}{2}\sqrt{\frac{2}{3}} (\tau_{11}-\tau_{22}\mp 2i\tau_{12}),
    \label{Eq:BipoSH_M2}
\end{align}
where $P_{\ell}(k)$ is the well-known Legendre coefficients for the Kaiser formula~\cite{Kaiser:1987},
\begin{align}
    P_{\ell=0}(k) =& \left( b_1^2 + \frac{2}{3}b_1 f+\frac{1}{5}f^2 \right) P_{\rm lin}(k)
    \\
    P_{\ell=2}(k) =& \left(\frac{4}{3}b_1f+\frac{4}{7}f^2 \right) P_{\rm lin}(k)
    \\
    P_{\ell=4}(k) =& \frac{8}{35}f^2 P_{\rm lin}(k).
\end{align}
We stress here that the isotropic signal is confined in the $L=0$ modes ($P^{00}_{\ell\ell}(k)$) which
do not suffer from the tidal mode
and the $L=2$ modes ($P^{2M}_{\ell\ell'}(k)$) successfully extracts the full five degrees of freedom of the super-sample tides. 
The explicit expressions of the BipoSH coefficients for the response to the super-sample density mode $D_{\ell}(k)$ and tidal mode $T_{\ell\ell'}(k)$ are given by
\begin{align}
    D_{0}(k) =& \left[-2b_1^3+\frac{47}{21}b_1^2 + 2b_1b_2 -\frac{1}{3}b_1^2 \frac{d\ln{P^L(k)}}{d\ln{k}}\right]
                +\left[\frac{26}{21}b_1 -b_1^2+\frac{2}{3}b_2 -\frac{1}{9}b_1(2+b_1) \frac{d \ln{P^L(k)}}{d \ln{k}} \right]f
    \nonumber\\
        &+\left[ \frac{31}{105} -\frac{8}{45}b_1 -\frac{1}{15}(1+2b_1)  \frac{d \ln{P^L(k)}}{d \ln{k}} \right]f^2
        +\left[-\frac{1}{105}-\frac{1}{21}\frac{d \ln{P^L(k)}}{d \ln{k}}  \right]f^3
        \\
    D_2(k) =& \left[ \frac{52}{21}b_1 -\frac{4}{3}b_1^2 + \frac{4}{3}b_2 -\frac{2}{9}b_1(2+b_1)\frac{d \ln{P^L(k)}}{d \ln{k}}  \right]f
    +\left[ \frac{124}{147}- \frac{8}{63}b_1-\frac{4}{21}(1+2b_1)\frac{d \ln{P^L(k)}}{d \ln{k}} \right]f^2
    \nonumber\\
        &+\left[\frac{4}{63}-\frac{10}{63}\frac{d \ln{P^L(k)}}{d \ln{k}} \right]f^3
        \\
    D_4(k) =& \left[\frac{248}{735}+\frac{32}{105}b_1 -\frac{8}{105}(1+2b_1)\frac{d \ln{P^L(k)}}{d \ln{k}} \right]f^2
                +\left[\frac{72}{385}-\frac{8}{77}\frac{d \ln{P^L(k)}}{d \ln{k}} \right]f^3
    \\
    D_6(k) =&\frac{16}{693}\left[4-\frac{d \ln{P^L(k)}}{d \ln{k}}  \right]f^3,
\end{align}
\begin{align}
    T_{20}(k) =& \left[ \frac{8}{7}b_1^2+2b_1b_{s^2} -b_1^2\frac{d \ln{P^L(k)}}{d \ln{k}}  \right]
                    +\left[ \frac{8}{7}b_1 +\frac{2}{3}b_{s^2} -\frac{1}{3}(2b_1+b_1^2)\frac{d \ln{P^L(k)}}{d \ln{k}} \right]f
    \nonumber\\
             &+\left[ \frac{16}{35} +\frac{4}{5}b_1 -\frac{1}{5}(1+2b_1)\frac{d \ln{P^L(k)}}{d \ln{k}}   \right]f^2
                +\left[ \frac{16}{35} -\frac{1}{7}\frac{d \ln{P^L(k)}}{d \ln{k}} \right]f^3
                \label{eq:T20}
    \\
    T_{02}(k) =& \left[ \frac{16}{35}b_1  - b_1^2 +\frac{4}{15}b_{s^2} -\frac{1}{15}b_1(4+5b_1)\frac{d \ln{P^L(k)}}{d \ln{k}} \right]f
                    +\left[ \frac{64}{245}  -\frac{8}{15}b_1-\frac{1}{35}(4+14b_1)\frac{d \ln{P^L(k)}}{d \ln{k}}  \right]f^2
    \nonumber\\
             &+\left[ \frac{13}{35} -\frac{1}{7}\frac{d \ln{P^L(k)}}{d \ln{k}} \right]f^3
             \label{eq:T02}
    \\
    T_{22}(k) =& \left[ \frac{32}{49}b_1 +\frac{8}{21}b_{s^2} -\frac{1}{21}b_1(8+7b_1) \frac{d \ln{P^L(k)}}{d \ln{k}}\right]f
                +\left[ \frac{128}{343}+ \frac{20}{147}b_1  -\frac{2}{49}(4+11b_1)\frac{d \ln{P^L(k)}}{d \ln{k}}   \right]f^2
    \nonumber\\
            &+\left[ \frac{76}{147}-\frac{25}{147}\frac{d \ln{P^L(k)}}{d \ln{k}} \right]f^3
    \\
    T_{42}(k) =& \left[ \frac{288}{245}b_1 +\frac{24}{35}b_{s^2} -\frac{24}{35}b_1\frac{d \ln{P^L(k)}}{d \ln{k}}\right]f
                    +\left[ \frac{1152}{1715}+\frac{144}{245}b_1 -\frac{72}{245}(1+b_1)\frac{d \ln{P^L(k)}}{d \ln{k}}   \right]f^2
    \nonumber\\
            &+\left[ \frac{144}{245}-\frac{8}{49}\frac{d \ln{P^L(k)}}{d \ln{k}} \right]f^3
    \\
    T_{24}(k) =&\left[\frac{256}{1715} -\frac{192}{245}b_1 - \frac{8}{245}(2+9b_1) \frac{d \ln{P^L(k)}}{d \ln{k}} \right]f^2
                    +\left[\frac{88}{245}-\frac{8}{49}\frac{d \ln{P^L(k)}}{d \ln{k}}   \right]f^3
    \\
    T_{44}(k) =&\left[\frac{512}{3773}+\frac{16}{49}b_1- \frac{8}{539}(4+11b_1) \frac{d \ln{P^L(k)}}{d \ln{k}} \right]f^2
                    +\left[\frac{2048}{5929}-\frac{600}{5929}\frac{d \ln{P^L(k)}}{d \ln{k}}   \right]f^3
    \\
    T_{64}(k) =&\left[\frac{128}{539}- \frac{8}{77} \frac{d \ln{P^L(k)}}{d \ln{k}} \right]f^2
                    +\left[\frac{160}{847}-\frac{40}{847}\frac{d \ln{P^L(k)}}{d \ln{k}}   \right]f^3
    \\
    T_{46}(k) =&\left[ \frac{72}{847}-\frac{40}{847}\frac{d \ln{P^L(k)}}{d \ln{k}} \right]f^3 
    \\
    T_{66}(k) =&\left[ \frac{32}{363}-\frac{8}{363}\frac{d \ln{P^L(k)}}{d \ln{k}} \right]f^3
                 \label{eq:T66}
\end{align}
Note that all other BipoSH coefficients are zero. Details of derivations of the above expressions are summarized in Appendix~\ref{app:sss_biposh}.

\begin{figure}[t]
	\centering
    \includegraphics[width=18cm]{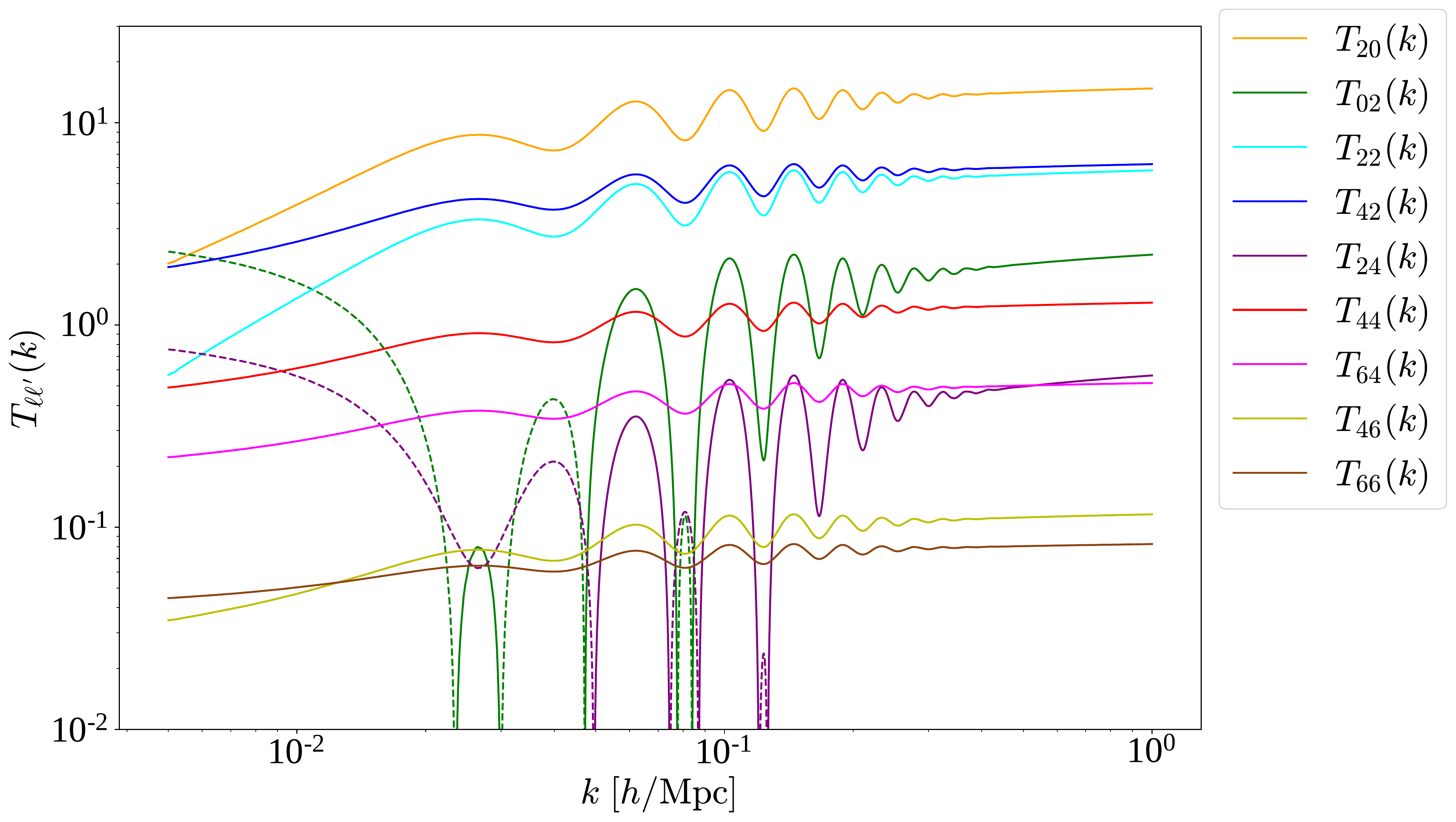}
	\caption{The response functions for the super-sample tides, $T_{\ell\ell'}(k)$, in terms of the BipoSH multipoles.
	Negative values are plotted with the dashed lines.
	Note that $T_{\ell\ell'}(k)$ is normalized by the matter power spectrum, i.e., $T_{\ell\ell'}(k) = \frac{\partial P^{20}_{\ell\ell'}}{\partial \tau_{33}}/P_m$.
	We use the following values; $f(z=0.8)=0.84$, $b_1=1.5$, $b_2 = 0.3$, and $b_{s^2}= -0.29$.}
	\label{fig:response}
\end{figure}

In Fig.~\ref{fig:response} we show the BipoSH multipole representation of the response functions for the super-sample tides,
$T_{\ell\ell'}(k)$,
which are normalized by the matter power spectrum, i.e, $\frac{\partial P^{20}_{\ell\ell'}}{\partial \tau_{33}}/P_m$.
We find that $T_{20}(k)$ has the highest magnitude, and $T_{02}(k)$, $T_{22}(k)$ and $T_{42}(k)$ have smaller and similar magnitudes.
In general, there is a hierarchy of magnitudes; $|P^{2M}_{20}|>|P^{2M}_{02}|\sim|P^{2M}_{22}|\sim|P^{2M}_{42}|>|P^{2M}_{24}|\sim|P^{2M}_{44}|\sim|P^{2M}_{64}|>|P^{2M}_{66}|$,
as reflected the power of the velocity.
Therefore $P^{2M}_{20}$ has a dominant contribution in the signal-to-noise ratio, similar to the usual Legendre multipole case in which $P_0$ is more dominant than $P_2$.
The wiggly feature in the response function results from the BAO phase shift described by the dilation terms as explained in Section \ref{subsec:effect_SS}.

\section{\label{sec:fisher}Fisher forecasts}
In this section, we study the degeneracies between the super-sample tidal modes and other cosmological anisotropies and the possibility to detect the super-sample tidal modes based on the Fisher information matrix formalism.
\subsection{Alcock-Paczynski effect}

In a spectroscopic survey, we infer the three-dimensional distances of galaxies
in the comoving coordinate from the angular and redshift separation $(\Delta \theta, \Delta z)$.
This transformation from $(\Delta \theta, \Delta z)$ to $(\Delta x_{\perp}, \Delta x_{\parallel})$ requires 
the angular diameter distance $D_A(z)$ and the Hubble expansion rate $H(z)$, which depend on the cosmological model we assume.

If the fiducial cosmological model we use differs from the underlying true cosmological model,
the relation between the comoving true wave vector and observed wave vector is given by
\begin{align}
    \vec{k}^{\rm true}_{\perp} = \frac{D_A^{\rm fid}}{D_A}\vec{k}^{\rm obs}_{\perp}, \hspace{1cm}
    \vec{k}^{\rm true}_{\parallel} = \frac{H}{H^{\rm fid}} \vec{k}^{\rm obs}_{\parallel}.
    \label{eq:AP_trans}
\end{align}
where the quantities with subscript "true" refer to the underlying true values and the quantities with "obs" are obtained from the assumed fiducial cosmological model.
Then the magnitude and line-of-sight (LOS) component of the comoving wave vector become
\begin{align}
    k^{\rm true} =& \sqrt{|\vec{k}^{\rm true}_{\perp}|^2 + |\vec{k}^{\rm true}_{\parallel}|^2 }
                 = \sqrt{ \left(\frac{D_A^{\rm fid}}{D_A}\right)^2|\vec{k}^{\rm obs}_{\perp}|^2 + \left(\frac{H}{H^{\rm fid}}\right)^2|\vec{k}^{\rm obs}_{\parallel}|^2} 
                 \nonumber\\
                 =& k^{\rm obs} \sqrt{\left(\frac{D_A^{\rm fid}}{D_A}\right)^2(1-\mu^2_{\rm obs}) +  \left(\frac{H}{H^{\rm fid}}\right)^2\mu^2_{\rm obs}} 
                 \nonumber\\
                 \equiv&  k^{\rm obs}\gamma(D_A^{\rm fid}, H^{\rm fid}, \mu_{\rm obs})
                 \\
    \mu^{\rm true} =& \frac{k^{\rm true}_{\parallel}}{k^{\rm true}}
                   = \frac{H}{H^{\rm fid}}\frac{1}{\gamma(D_A^{\rm fid}, H^{\rm fid}, \mu_{\rm obs})}\frac{k^{\rm obs}_{\parallel}}{k^{\rm obs}}
                   \nonumber\\
                   =& \frac{H}{\gamma H^{\rm fid}}\mu^{\rm obs}.
\end{align}
This leads to the observed power spectrum in redshift space,
\begin{align}
    P_{\rm obs}(k^{\rm obs}, \mu^{\rm obs}  ) = \frac{H}{H^{\rm fid}} \left(\frac{D^{\rm fid}_A}{D_A} \right)^2 P(k^{\rm true} ,\mu^{\rm true}),
    \label{eq:AP_power}
\end{align}
which means that there appear the higher-order multipoles than $\ell=4$ if the fiducial cosmology does not match the underlying true cosmology.
This geometrical distortion is called the Alcock-Paczynski effect.
In general, the AP effect makes the galaxy clustering anisotropic even in the absence of the RSD.
Notice that, however, the generated anisotropic signals are confined into the LOS direction and
there is no anisotropic distortion in the plane perpendicular to the LOS due to the AP effect.
In other words, the AP effect leaves the redshift-space power spectrum two-dimensional one, 
which still respects the three-dimensionally rotational symmetry around the observer.
In terms of the BipoSH, the information content carried by the AP distortion can be completely captured by the $L=0$ multipoles.

\subsection{Fisher information matrix}
We employ the Fisher matrix formalism in order to assess the correlations between the anisotropic signals that appear in the observed power spectrum of galaxies.  
In terms of the reduced BipoSH coefficient $P^{LM}_{\ell_1\ell_2}(k)$, the Fisher matrix is written as 
\begin{align}
    F_{\alpha \beta} &= -\left\langle \frac{\partial \log{L}}{\partial \theta_{\alpha} \partial \theta_{\beta} } \right\rangle  
    \nonumber \\
        &= \sum_{kk'} \sum_{\ell_1\ell'_1\ell_2\ell'_2}\sum_{LL'MM'} \frac{\partial [P^{LM}_{\ell_1 \ell_2}(k)]^*}{\partial \theta_\alpha}
        {\rm Cov}^{-1}\left[ [P^{LM}_{\ell_1 \ell_2}(k)]^*  ,P^{L'M'}_{\ell'_1 \ell'_2}(k') \right]
        \frac{\partial P^{L'M'}_{\ell'_1 \ell'_2}(k')}{\partial \theta_\beta} \label{eq:Fisher_matrix},
\end{align}
where $L$ is the likelihood and $\theta_{\alpha}$ is the $\alpha$-th parameter of interest.
The Cram\'er-Rao bound states that the minimum possible errors on parameter $\alpha$, marginalized over all other parameters, are given by 
the square root of the diagonal components of the inverse of the Fisher matrix as
\begin{align}
   \Delta \theta_{\alpha} \ge \sqrt{(F^{-1})_{\alpha\alpha}},
\end{align}
while the unmarginalized ones are given by $\Delta \theta_{\alpha} = 1/\sqrt{F_{\alpha\alpha}}$.
The cross-correlation coefficients $c_{\alpha\beta}$ are defined through
\begin{align}
    c_{\alpha\beta} = \frac{(F^{-1})_{\alpha\beta}}{\sqrt{ (F^{-1})_{\alpha\alpha} (F^{-1})_{\beta\beta} }}.
\end{align}

To give an expression of the covariance matrix for the reduced BipoSH coefficients $P^{LM}_{\ell_1 \ell_2}(k)$,
we start from the covariance for the 3D redshift-space power spectrum of galaxies $P^s({\bf k}; \hat{n})$ in Gaussian limit,
\begin{align}
    \textrm{Cov}\left[P^s({\bf k}; \hat{n}), P^s({\bf k}'; \hat{n'})\right]
    = 4\pi \frac{\delta^K_{k,k'}}{N_{k}} 
    \left[\sum_J P^{(O)}_J(k) \mathcal{L}_J(\hat{k}\cdot\hat{n}) \right]^2
    \left[ \delta^{(2)}(\hat{k}+\hat{k}') + \delta^{(2)}(\hat{k}-\hat{k}')  \right]
    4\pi \delta^{(2)}(\hat{n}-\hat{n}'),
\end{align}
where $N_k = 4\pi k^2\Delta k V/(2\pi)^3$ is the number of modes with survey volume $V$ and the interval between each Fourier mode $\Delta k$
and $\mathcal{L}_{J}(x)$ is the Legendre polynomial.
The Legendre coefficients with subscript $(O)$ denote $P_0^{(O)}(k) = P_0(k)+1/\bar{n}_g$, $P_2^{(O)}(k) = P_{2}(k)$, $P_4^{(O)}(k) = P_{4}(k)$ and $P_1^{(O)}(k) = P_3^{(O)}(k) = P_{J\geq 5}^{(O)}(k) = 0$
with $\bar{n}_g$ being the local number density of galaxies.
By making use of the formulas in Appendix \ref{app:Identitis}, the covariance above leads to the following expression of the covariance 
for the reduced BipoSH coefficients~\cite{Shiraishi_etal:2016},
\begin{align}
    \textrm{Cov}\left[ [P^{LM}_{\ell_1 \ell_2}(k)]^*, P^{L'M'}_{\ell'_1 \ell'_2}(k') \right] 
    = \delta^K_{L,L'} \delta^K_{M,M'} \frac{\delta^K_{k,k'}}{N_k}\Theta^L_{\ell_1,\ell_2, \ell'_1, \ell'_2}(k),
    \label{eq:covariance_biposh}
\end{align}
where
\begin{align}
    \Theta^L_{\ell_1, \ell_2, \ell'_1, \ell'_2}(k)
    =& (2\ell_1+1)(2\ell_2+1)(2\ell'_1+1)(2\ell'_2+1)(2L+1)(-1)^{\ell_1}\left[1+(-1)^{\ell'_1}\right]
    H_{\ell_1\ell_2 L}H_{\ell'_1\ell'_2 L}
    \sum_{JJ'}P^{(O)}_J(k) P_{J'}^{(O)}(k)
    \nonumber \\
    &\sum_{L_1 L_2}(2L_1+1)(2L_2+1)
    H_{\ell_1 J L_1}H_{\ell_2 J L_2} H_{\ell'_1 J' L_1}H_{\ell'_1 J' L_2}
    \begin{Bmatrix}
    L & L_1 & L_2 \\
    J & \ell_2 & \ell_1
    \end{Bmatrix}
    \begin{Bmatrix}
    L & L_1 & L_2 \\
    J' & \ell'_2 & \ell'_1
    \end{Bmatrix},
\end{align}
with
$\left\{\begin{smallmatrix}
L_1 & L_2 & L_3 \\
L_4 & L_5 & L_6
\end{smallmatrix}\right\}
$
being the Wigner 6$j$ symbol.
The covariance matrix is therefore block-diagonalized for $L$, $M$ and $k$.
Further, the reality of the covariance means the matrix is also block-diagonalized for real part and imaginary part.

\subsection{Results}
In this paper, we use the following cosmological parameters that are consistent with the Planck 2018 results~\cite{Planck2018_cosmo}:
$h=0.6766$,
$\Omega_c h^2= 0.1193$,
$\Omega_b h^2= 0.0224$,
$A_s= 2.105\times10^{-9}$ and
$n_s= 0.9665$.
We compute the Fisher matrix Eq.~(\ref{eq:Fisher_matrix}) for the following parameter set,
\begin{align}
    \theta_\alpha = \left\{b_1, b_2, b_{s^2}, f, D_A(z), H(z), \delta_{\rm b}, \tau_{33}, \tau_{11}, \tau_{12},\tau_{13}, \tau_{23}  \right\}.
\end{align}
As a working example, we assume a SPHEREx-like survey where we set
the fiducial values of 
the central redshift $z=0.8$, the comoving survey volume $V_{\rm survey}= 4.0 \ ({\rm Gpc}/h)^3$, 
the mean number density of galaxies $\bar{n}_g = 4.0\times 10^{-3}\ (h/{\rm Gpc})^3$,
the linear bias $b_1 = 1.5$, the quadratic bias $b_2 = 0.3$, the tidal bias $b_{s^2} = -\frac{4}{7}(b_1-1)=-0.29$,
and the linear growth rate $f(z=0.8)=0.84$~\cite{Spherex_wp}.
Here we consider a single redshift slice for simplicity, but our results can be trivially extended to include multiple redshift slice.
Notice however that because the super-sample modes $\delta_{\rm b}(z_i)$ and $\tau_{ij}(z_i)$ depend on the specific survey region one should treat the $\delta_{\rm b}(z_i)$ and $\tau_{ij}(z_i)$
as independent variables for each redshift slice, unless one considers the super-sample modes that straddle multiple survey regions.

\begin{figure}[b]
	\centering
    \includegraphics[width=12cm]{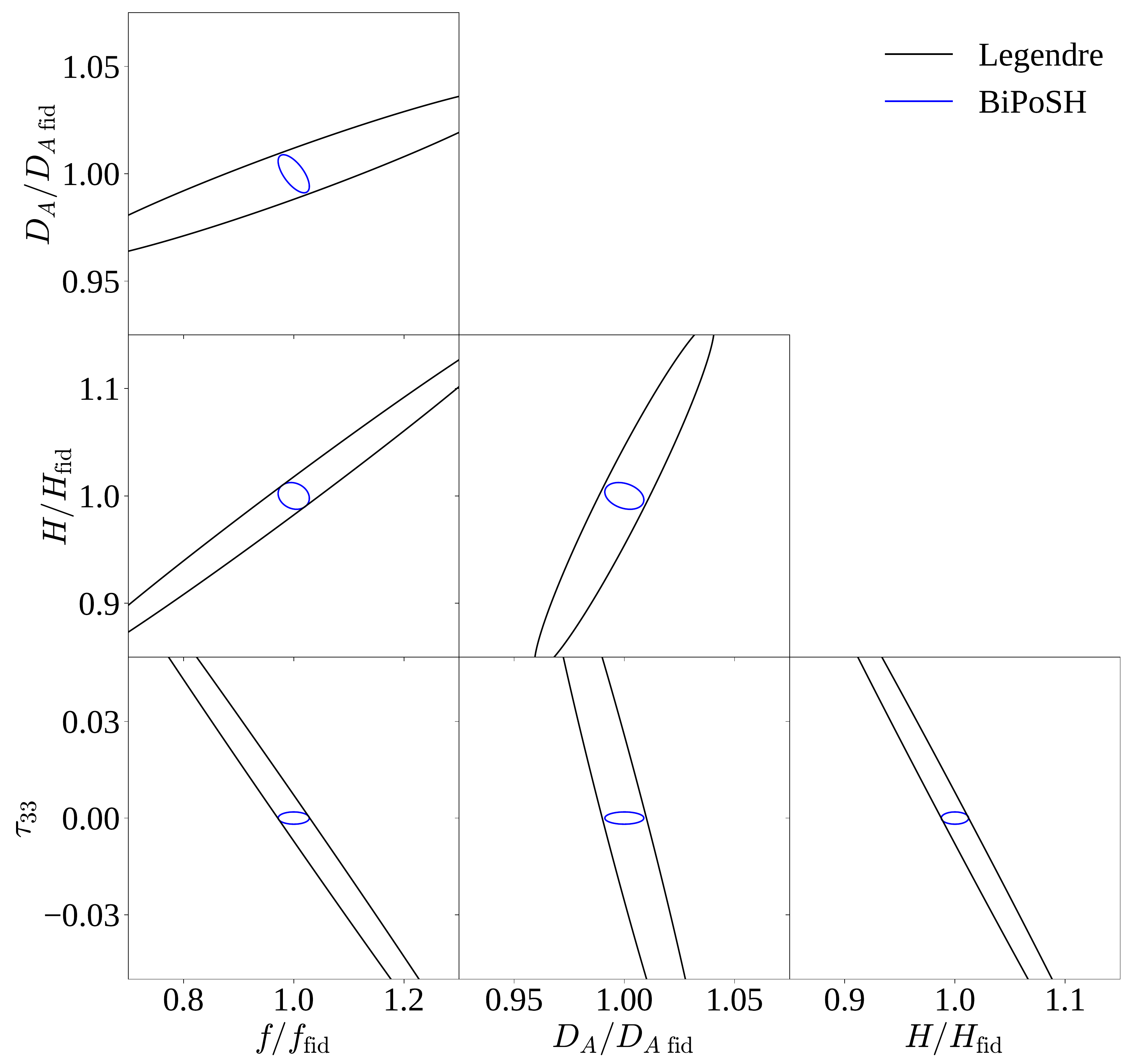}
	\caption{1$\sigma$ (68\%) error contour for joint $\tau_{33}$ and cosmological distortion parameter, $f, D_A$ and $H$, estimation with the maximum wavenumber $k_{\rm max}=0.2\ h/{\rm Mpc}$. 
	The inner blue curves in each panel show the results when employing the BipoSH expansion, which carries the full information of the three-dimensional power spectrum. The outer black curves in each panel correspond to the results when using the Legendre expansion, which contains only two-dimensional information of the power spectrum. 
	}
	\label{fig:fisher_ellipse_spherex}
\end{figure}

The two comments are in order.
First, since the higher-order biases, $b_2$ and $b_{s^2}$, are only in the response functions at tree level calculation, then
information is not sufficient to determine these higher-order biases.
Therefore we employ 3$\sigma$ Gaussian priors for $b_2$ and $b_{s^2}$ with $\sigma_{b_2} = \sigma_{b_{s^2}} = 1$
in order to make the Fisher matrix invertible.
Second, because the isotropic component of the super-sample modes $\delta_{\rm b}$ is to degenerate with the linear bias $b_1$ in spectroscopic surveys \cite{Li_etal:2014b,Chiang_Slosar:2018},
we also add a 3$\sigma$ Gaussian prior to $\delta_{\rm b}$ with $\sigma_{\rm b}$ computed from Eq.~(\ref{eq:variance_delta}) and focus on investigating
the degeneracies and detectability of $\tau_{ij}$ in this paper.
Note however that it is possible to constrain $\delta_{\rm b}$ in lensing surveys where the global mean density is relevant~\cite{Li_etal:2014b}.

Fig.~\ref{fig:fisher_ellipse_spherex} shows the marginalized 68\% error contours for the anisotropic signals $\{f, D_A, H, \tau_{33}\}$ 
in each of two-dimensional sub-space when adopting the minimum wavenumber $k_{\rm min}=5.0\times 10^{-3}\ h/{\rm Mpc}$, which is larger than the fundamental modes $k_F\sim 2\pi/V^{1/3}$, and the maximum wavenumber $k_{\rm max} = 0.2\ h/{\rm Mpc}$. 
We only present the results of $\tau_{33}$
because the results are identical for $\tau_{11}, \tau_{12}, \tau_{13}$, and $\tau_{23}$ in the BipoSH expansion.
For comparison of the BipoSH expansion with the Legendre expansion, both expansion scheme cases are plotted.
We use Eq.~(\ref{eq:legendre_0})-(\ref{eq:legendre_6}) as the Legendre multipoles.
Fig.~\ref{fig:fisher_ellipse_spherex} clearly demonstrates the power of the BipoSH expansion compared with the Legendre expansion.
Using the Legendre coefficients, the super-sample tidal mode significantly degrades the constraints on other parameters
while in the BipoSH expansion the super-sample tidal modes have little impact on the estimation of other parameters.
More quantitatively, the absolute values of the cross-correlation coefficients between the $\tau_{33}$ and other parameters $\beta=\{b_1, f, D_A, H\}$, $c_{\tau\beta}$, are less than $\mathcal{O}(0.1)$.

\begin{figure}[t]
	\centering
    \includegraphics[width=15cm]{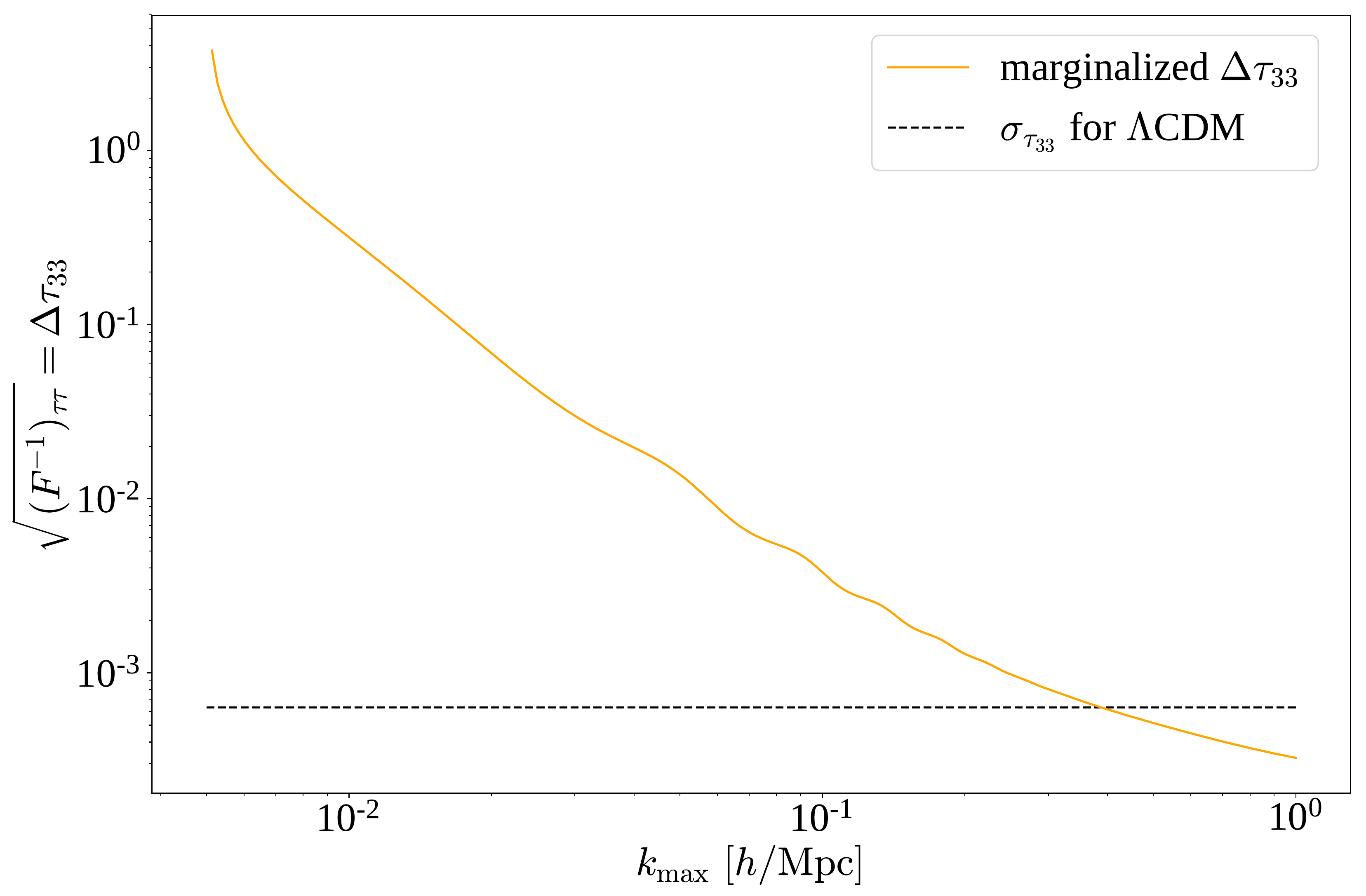}
	\caption{1$\sigma$ (68\%) error for $\tau_{33}$ as a function of the maximum wavenumber $k_{\rm max}$.
	The orange curve corresponds to the 1$\sigma$ constraint on $\tau_{33}$ when marginalized over other parameters, $\sqrt{(F^{-1})_{\tau\tau}}$.
	The horizontal dashed line represents the rms value, $\sigma_{\tau_{33}}$, expected for $\Lambda$CDM cosmology and the survey volume, which is calculated from Eq. (\ref{eq:variance_tau}).}
	\label{fig:fisher_tau_spherex}
\end{figure}

Mathematically this is a consequence of the following facts.
First, in the BipoSH expansion $f, D_A$ and $H$ are confined into the isotropic $L=0$ BipoSH multipoles, $P^{00}_{\ell\ell}$,
wheres the tidal signals $\tau_{ij}$ are confined into the $L=2$ BipoSH multipoles, $P^{2M}_{\ell\ell'}$.
Second, the covariance for the BipoSH coefficients (Eq. (\ref{eq:covariance_biposh})) is block diagonal for $L$.
Then $f, D_A$ and $H$ are constrained mainly from $P^{00}_{\ell\ell}(k)$, to which $\tau_{ij}$ do not contribute.
On the other hand, in the Legendre expansion $f, D_A, H$ and $\tau_{33}$ all appear in the same multipoles, $P_{\ell}(k)$.
Accordingly, changing the AP parameters, $D_A$ and $H$, leads to both the growth-like and dilation-like effect on the Legendre multipoles \cite{Padmanabhan_etal:2008},
and changing the RSD parameter $f$ mimics the growth effect due to the super-sample mode.
Hence, these parameters degenerate with each other in the Legendre expansion.

The little correlation between $\tau_{ij}$ and other distortion parameters suggests that the large-scale tides $\tau_{ij}$ can be measured
from the galaxy redshift-space power spectrum if using the BipoSH expansion,
and therefore it is worthwhile to explore the possibility of whether a spectroscopic survey can detect $\tau_{ij}$
when including higher $k_{\rm max}$.
Fig.~\ref{fig:fisher_tau_spherex} shows the 1$\sigma$ constraint on $\tau_{33}$ as a function of $k_{\rm max}$.
Again, the constraints on other super-sample tidal modes are equivalent,
so we only present the $\tau_{33}$ estimation.
We find that the BipoSH expansion enable\Mare{s} us to determine $\tau_{33}$ with an accuracy better than the rms of $\tau_{33}$ expected for $\Lambda$CDM model,
simultaneously measuring the RSD and AP distortion,
if $P^{20}_{\ell\ell'}(k)$ is included up to $k_{\rm max} \gtrsim 0.3\ h/{\rm Mpc}$.

Although the inclusion of high $k_{\rm max}$, in general, requires to accurately model the nonlinear effect, 
the nonlinear evolution cannot generate the azimuthal asymmetry about the LOS, and therefore $P^{2M}_{\ell\ell'}(k)$.
Hence, the fact that the appearance of $P^{2M}_{\ell\ell'}(k)$ is a distinctive feature of $\tau_{ij}$
allow us to use high $k_{\rm max}$ value. 
In practice, the highest $k_{\rm max}$ value we can use is limited by our knowledge of the response of the redshift-space power spectrum to the large-scale tides, 
$\partial P({\bf k},\hat{n})/\partial \tau_{ij}$,
in the nonlinear regime where the perturbation theory breaks down.
To know the form of $\partial P({\bf k},\hat{n})/\partial \tau_{ij}$ in the nonlinear regime, we need to run the $N$-body simulation with the large-scale tidal field~\cite{Schmidt_etal:2018},
which is the similar method used for estimating the response for the mean density modulation, $\partial P(k,\mu)/\partial \delta_{\rm b}$, 
called as the ``separate universe simulation"~\cite{Sirko:2005,Li_etal:2014a,Li_etal:2014b,Wagner_etal:2014,Baldauf_etal:2015,Jamieson_LoVerde:2018}.
This is beyond the scope of this paper, so here we simply assume that the response function derived from the tree-level calculation of perturbation theory holds in the nonlinear regime,
and this approximation could be suitable up to $k=0.3 \ h/{\rm Mpc}$~\cite{Schmidt_etal:2018}.

\section{\label{sec:conclusion}Conclusions}
In this paper, we, for the first time, have used 
the BipoSH decomposition formalism of the galaxy power spectrum
to assess how the super-sample modes, $\delta_{\rm b}$ and $\tau_{ij}$, have the influence on 
the measurements of other cosmological distortions including the RSD and AP effects
in an all-sky galaxy spectroscopic survey like the SPHEREx survey.

The super-sample tidal components, $\tau_{ij}$, degenerate with cosmological parameters
of our interest such as the growth rate function $f$, the Hubble parameter $H$ and the angular diameter distance $D_{\rm A}$~\cite{Akitsu_Takada:2017}.
To break the parameter degeneracy, it is essential to notice the fact that $\tau_{ij}$ breaks the statistical isotropy of the observed galaxy power spectrum.
The BipoSH formalism characterizes statistical anisotropic signals via a multipole index $L$ in the BipoSH basis $\{Y_{\ell}(\hat{k})\otimes Y_{\ell'}(\hat{n}) \}_{LM}$,
and non-zero $L$ modes mean the presence of statistical anisotropy; in other words, 
we can single out only the statistical anisotropic signal by measuring the $L\neq 0$ mode.
The super-sample tidal components result in the $L=2$ mode, and we have shown the explicit expressions of all non-vanishing BipoSH components in Eqs.~(\ref{Eq:BipoSH_M0})-(\ref{Eq:BipoSH_M2}) and Eqs.~(\ref{eq:T20})-(\ref{eq:T66}).

To see how the BipoSH formalism works well in order to break the parameter degeneracy, 
we have performed the Fisher matrix computation.
Assuming the SPHEREx-like survey, we have found that $\tau_{ij}$ can be constrained 
with $\Delta\tau_{ij} \lesssim \mathcal{O} (10^{-3})$ 
through the $L=2$ mode measurement,
and therefore, it has little impact on the estimation of the RSD and AP effect, whose signals are confined to the $L=0$ mode.
In other words, we could obtain information on super-survey modes beyond a finite survey region through the well-constrained $\tau_{ij}$. 
This enables unbiased estimations on $f$, $H$ and $D_A$.

Finally, we summarize some possible applications of the method presented in this paper.
The first one is to confirm the matter-radiation equality bump in the power spectrum 
that is predicted in the linear cosmological perturbation theory.
Assuming that there is no decrease in the power spectrum at low $k$ unlike the $\Lambda$CDM model, 
the super-sample tides predicted from Eq.~(\ref{eq:variance_tau}) should become larger 
than the $\Lambda$CDM prediction.
To put it the other way, a non-detection of such large tides can rule out the model that has no matter-radiation equality bump in the power spectrum.
Related to this direction, the second one lies in
exploring the large-scale anomaly such as the super-curvature fluctuation~\cite{Nan_etal:2019} and quintessential isocurvature~\cite{Jamieson_LoVerde:2018}.
To do so, we need to calibrate the response of small-scale perturbation to the large-scale fluctuation depending on each model.
The final one is related to constraining anisotropic inflation models. Signals due to such models are also present in the $L=2$ mode \cite{Shiraishi_etal:2016,Sugiyama_etal:2017,Bartolo:2017sbu} and hence may be biased by the super-sample tides. As has been done in this paper, the distinguishability should be examined for interpreting observational constraints precisely. We leave these for future works.

\section*{\label{sec:Acknowledgments}Acknowledgments}
We thank to Masahiro Takada, Chi-Ting Chiang, and An\v{z}e Sloar for helpful discussions.
KA is supported by the JSPS Research Fellowship for Young Scientists and Advanced Leading Graduate Course for Photon Science at the University of Tokyo.
NSS acknowledges financial support from Grant-in-Aid for JSPS Fellows (Nos. 28-1890)
and from JSPS KAKENHI Grant Number 19K14703. 
MS is supported by JSPS Grant-in-Aid for Research Activity Start-up Grant Number 17H07319 and JSPS Grant-in-Aid for Early-Career Scientists Grant Number 19K14718.

\appendix

\section{\label{app:relation}Relationship between the BipoSH and the other expansion schemes}
In this appendix, we give the connection between our formalism and other schemes to decompose anisotropic signals.
\subsection{Relationship between the BipoSH and the Legendre expansion}
The relationship between the BoPoSH and the usual Legendre expansion with the line-of-sight $\hat{n}$ setting $\hat{z}$ is
\begin{align}
    P_{\ell}(k) =& (2\ell+1)\int \frac{d^2\hat{k}}{4\pi} \int\frac{ d^2\hat{n}}{4\pi} ~ P^s({\bf k}, \hat{n}; \delta_{\rm b}, \tau_{ij}) 
                    \mathcal{L}_\ell(\hat{k}\cdot\hat{n})  (4\pi) \delta^{(2)}_D(\hat{n}-\hat{z}) 
        \nonumber\\
                =& \sum_{LM\ell'}\pi^{LM}_{\ell\ell'}(k; \delta_{\rm b}, \tau_{ij})
                     (-1)^{\ell-\ell'+M}
                    \sqrt{\frac{(2\ell+1)(2\ell'+1)(2L+1)}{(4\pi)^2}} 
                    \begin{pmatrix}
                    \ell & \ell' & L \\
                    0 & 0 & -M
                    \end{pmatrix}
        \nonumber\\
        =&  \sum_{L\ell'}P^{L0}_{\ell\ell'}(k; \delta_{\rm b}, \tau_{ij})
\end{align}
where we used $Y_{\ell m}(\hat{z}) = \sqrt{(2\ell+1)/(4\pi)}\delta_{m0}$. 
Notice that the summation of $L$ and $\ell'$ is limited by the Wigner 3$j$ symbol 
$\left(\begin{smallmatrix}
\ell & \ell' & L \\
0 & 0 & 0 
\end{smallmatrix}\right)$.
For instance,
\begin{align}
    P_{\ell=0}(k) =& P^{00}_{00}(k) + P^{20}_{02}(k) \label{eq:legendre_0}  \\
    P_{\ell=2}(k) =& P^{00}_{22}(k) + P^{20}_{20}(k) + P^{20}_{22}(k) + P^{20}_{24}(k) \label{eq:legendre_2} \\
    P_{\ell=4}(k) =& P^{00}_{44}(k) + P^{20}_{42}(k) + P^{20}_{44}(k) + P^{20}_{46}(k) \label{eq:legendre_4} \\
    P_{\ell=6}(k) =& P^{20}_{64}(k) + P^{20}_{66}(k). \label{eq:legendre_6} \\
\end{align}
This reproduces the result presented in Appendix of Ref.\citep{Akitsu_Takada:2017}.

\subsection{Relationship between the BipoSH and the single spherical harmonic expansion}
In Ref. \citep{Chiang_Slosar:2018}, the authors defined the spherical multipole expansion,
\begin{align}
P_{gg}({\bf k}; \hat{n}\parallel \hat{z}) &= \sum_{\ell m} P_{gg;\ell m}(k) Y_{\ell m}(\hat{k})\\
P_{gg;\ell m}(k) &= \int d^2\hat{k}~ P_{gg}({\bf k}; \hat{n}\parallel \hat{z})Y^*_{\ell m}(\hat{k}).
\end{align}
The transformation from the BipoSH expansion to the spherical multipole expansion is 

\begin{align}
    P_{gg,\ell m}(k)
        =& \int \frac{d^2\hat{k}}{4\pi}\int \frac{d^2\hat{n}}{4\pi}~ P^s({\bf k}, \hat{n}; \delta_{\rm b}, \tau_{ij}) Y^*_{\ell m}(\hat{k}) (4\pi) \delta^{(2)}_D(\hat{n}-\hat{z}) 
        \nonumber\\
        =& \sum_{L\ell'}
            \pi^{Lm}_{\ell\ell'}(k; \delta_{\rm b}, \tau_{ij})
             (-1)^{\ell-\ell'+m}\sqrt{\frac{(2\ell'+1)(2L+1)}{(4\pi)^3}}
            \begin{pmatrix}
            \ell & \ell' & L \\
            m & 0 & -m
            \end{pmatrix}.
\end{align}

\section{\label{app:sss_biposh}BipoSH coefficients for the super-samle modes}
In this appendix, we provide how to derive the explicit expressions of the BipoSH coefficients for the responses to the super-sample modes.

\subsection{\label{app:cal_biposh}Calculation of the BipoSH coefficients}
In this subsection, we provide a derivation of each BipoSH coefficient for each type of the anisotropic terms.
\subsubsection{BipoSH coefficients from $ (\hat{k}\cdot\hat{n})^{\lambda}$ terms}
For later convenience, we decompose $(\hat{k}\cdot\hat{n})^{\lambda}$ into the spherical harmonic basis as
\begin{align}
(\hat{k}\cdot\hat{n})^{\lambda}
=\sum_{n=0}^{\lambda}A_{n\lambda}\mathcal{L}_n(\hat{k}\cdot\hat{n})
=\sum_{n=0}^{\lambda}\frac{4\pi A_{n\lambda}}{2n+1}\sum_{\nu}Y_{n\nu}(\hat{k})Y^*_{n\nu}(\hat{n}),
\end{align}
where 
\begin{align}
A_{n\lambda}&=\frac{2n+1}{2}\int_{-1}^{1}d\mu~ \mu^\lambda \mathcal{L}_n(\mu)\nonumber\\
&=\frac{2n+1}{2}[(-1)^{n+\lambda}+1]\int_{0}^{1}d\mu~ \mu^\lambda \mathcal{L}_n(\mu)\nonumber\\
&=\frac{2n+1}{2}\frac{(-1)^{n+\lambda}+1}{2^n}\frac{\Gamma(\lambda+1)\Gamma(\frac{\lambda-n+3}{2})}{\Gamma(\lambda-n+2)\Gamma(\frac{\lambda+n+3}{2})}
\Theta_{\lambda\geq n},
\end{align}
with $\Theta_{\lambda\geq n}\equiv \left\{  \begin{aligned}
&1 : \lambda \geq n \\
&0 : \lambda < n
\end{aligned}\right.$ being the step function.
Then, the BipoSH coefficients from $ (\hat{k}\cdot\hat{n})^{\lambda}$ terms are calculated as 
\begin{align}
\pi_{\ell \ell'}^{LM}(k)
&=\int  d^2\hat{k}\int d^2\hat{n}~
(\hat{k}\cdot\hat{n})^\lambda 
\sum_{mm'}\mathcal{C}^{LM}_{\ell m \ell' m'}Y^*_{\ell m}(\hat{k})Y^*_{\ell' m'}(\hat{n})\nonumber\\
&=\frac{4\pi A_{\ell \lambda}}{2\ell+1}
\delta^K_{\ell\ell'}
\sqrt{2L+1}(-1)^{\ell}
\sum_m 
(-1)^{\ell-m}
\begin{pmatrix}
\ell & \ell & L \\
m & -m & M
\end{pmatrix}
\nonumber\\
&=\frac{4\pi A_{\ell \lambda}}{\sqrt{2\ell+1}}
(-1)^{\ell}
\delta^K_{L0}\delta^K_{M0}\delta^K_{\ell\ell'}.
\end{align}

\subsubsection{BipoSH coefficients from $ (\hat{k}\cdot\hat{n})^{\lambda}\tau_{ij}\hat{k}_i\hat{k}_j$  terms}
\begin{align}
(\hat{k}\cdot\hat{n})^\lambda \tau_{ij} \hat{k}_i \hat{k}_j
&=\sum_{n=0}^\lambda \frac{4\pi A_{n\lambda}}{2n+1}
\sum_{\nu} Y^*_{n\nu}(\hat{k})Y_{n\nu}(\hat{n})
\left(\tau_{ij} \sum_{m_i m_j} \alpha^{m_i}_i Y_{1m_i}(\hat{k}) \alpha^{m_j}_j Y_{1m_j}(\hat{k}) \right) 
\nonumber\\
&=\sum_{n=0}^\lambda \frac{4\pi A_{n\lambda}}{2n+1}
\sum_{\nu} Y^*_{n\nu}(\hat{k})Y_{n\nu}(\hat{n}) \tau_{ij}
\sum_{m_i m_j} \alpha^{m_i}_i \alpha^{m_j}_j \sum_{\ell_3 m_3}
h_{11\ell_3} Y^*_{\ell_3 m_3}(\hat{k})
\begin{pmatrix}
1 & 1 & \ell_3 \\
m_i & m_j & m_3
\end{pmatrix}\nonumber\\
&=\sum_{n=0}^\lambda \frac{4\pi A_{n\lambda}}{2n+1}\tau_{ij}
\sum_{\nu}Y_{n\nu}(\hat{n})
\sum_{m_i m_j} \alpha^{m_i}_i \alpha^{m_j}_j \sum_{\ell_3 m_3}
h_{11\ell_3}
\begin{pmatrix}
1 & 1 & \ell_3 \\
m_i & m_j & m_3
\end{pmatrix}
\sum_{\ell''m''}
h_{n \ell_3 \ell''}
\begin{pmatrix}
n & \ell_3 & \ell'' \\
\nu & m_3 & m''
\end{pmatrix}
Y_{\ell'' m''}(\hat{k}) ,
\end{align}
where $h_{l_1 l_2 l_3} \equiv \sqrt{\frac{(2 l_1 + 1)(2 l_2 + 1)(2 l_3 + 1)}{4 \pi}}
\left(\begin{smallmatrix}
  l_1 & l_2 & l_3 \\
  0 & 0 & 0
\end{smallmatrix}\right)$. Then, we have
\begin{align}
\pi_{\ell \ell'}^{LM}(k)
&=\int d^2\hat{k}\int d^2\hat{n}~
(\hat{k}\cdot\hat{n})^\lambda \tau_{ij} \hat{k}_i\hat{k}_j
\sum_{mm'}\mathcal{C}^{LM}_{\ell m \ell' m'}Y^*_{\ell m}(\hat{k})Y^*_{\ell' m'}(\hat{n})\nonumber\\
&=\frac{4\pi A_{\ell' \lambda}}{2\ell'+1}\tau_{ij}
\sum_{m_i m_j}\alpha^{m_i}_i \alpha^{m_j}_j \sum_{\ell_3 m_3}
h_{11\ell_3}
\begin{pmatrix}
1 & 1 & \ell_3 \\
m_i & m_j & m_3
\end{pmatrix}
\sum_{mm'}\mathcal{C}^{LM}_{\ell m \ell' m'}
h_{\ell' \ell_3 \ell}
\begin{pmatrix}
\ell' & \ell_3 & \ell \\
m' & m_3 & m
\end{pmatrix}
\nonumber\\
&=\frac{4\pi A_{\ell' \lambda}}{2\ell'+1}\tau_{ij}
\sum_{m_i m_j}\alpha^{m_i}_i \alpha^{m_j}_j
\begin{pmatrix}
1 & 1 & L \\
m_i & m_j & -M
\end{pmatrix}
\frac{(-1)^{\ell-\ell'+M}}{\sqrt{2L+1}}
h_{11L} h_{\ell\ell'L}.
\end{align}

\subsubsection{BipoSH coefficients from $ (\hat{k}\cdot\hat{n})^{\lambda}\tau_{ij}\hat{n}_i\hat{n}_j$  terms}

\begin{align}
(\hat{k}\cdot\hat{n})^\lambda \tau_{ij} \hat{n}_i \hat{n}_j
&=\sum_{n=0}^\lambda \frac{4\pi A_{n\lambda}}{2n+1}
\sum_{\nu} Y_{n\nu}(\hat{k})Y^*_{n\nu}(\hat{n})
\left(\tau_{ij} \sum_{m_i m_j} \alpha^{m_i}_i Y_{1m_i}(\hat{n}) \alpha^{m_j}_j Y_{1m_j}(\hat{n}) \right) 
\nonumber\\
&=\sum_{n=0}^\lambda \frac{4\pi A_{n\lambda}}{2n+1}
\sum_{\nu} Y_{n\nu}(\hat{k})Y^*_{n\nu}(\hat{n}) \tau_{ij}
\sum_{m_i m_j} \alpha^{m_i}_i \alpha^{m_j}_j \sum_{\ell_3 m_3}
h_{11\ell_3} Y^*_{\ell_3 m_3}(\hat{n})
\begin{pmatrix}
1 & 1 & \ell_3 \\
m_i & m_j & m_3
\end{pmatrix}\nonumber\\
&=\sum_{n=0}^\lambda \frac{4\pi A_{n\lambda}}{2n+1}\tau_{ij}
\sum_{\nu}Y_{n\nu}(\hat{k})
\sum_{m_i m_j} \alpha^{m_i}_i \alpha^{m_j}_j \sum_{\ell_3 m_3}
h_{11\ell_3}
\begin{pmatrix}
1 & 1 & \ell_3 \\
m_i & m_j & m_3
\end{pmatrix}
\sum_{\ell''m''}
h_{n \ell_3 \ell''}
\begin{pmatrix}
n & \ell_3 & \ell'' \\
\nu & m_3 & m''
\end{pmatrix}
Y_{\ell'' m''}(\hat{n}).
\end{align}
Then, we have
\begin{align}
\pi_{\ell \ell'}^{LM}(k)
&=\int d^2\hat{k}\int d^2\hat{n}~
(\hat{k}\cdot\hat{n})^\lambda \tau_{ij} \hat{n}_i\hat{n}_j
\sum_{mm'}\mathcal{C}^{LM}_{\ell m \ell' m'}Y^*_{\ell m}(\hat{k})Y^*_{\ell' m'}(\hat{n})\nonumber\\
&=\frac{4\pi A_{\ell \lambda}}{2\ell+1}\tau_{ij}
\sum_{m_i m_j}\alpha^{m_i}_i \alpha^{m_j}_j \sum_{\ell_3 m_3}
h_{11\ell_3}
\begin{pmatrix}
1 & 1 & \ell_3 \\
m_i & m_j & m_3
\end{pmatrix}
\sum_{mm'}\mathcal{C}^{LM}_{\ell m \ell' m'}
h_{\ell \ell_3 \ell'}
\begin{pmatrix}
\ell & \ell_3 & \ell' \\
m & m_3 & m'
\end{pmatrix}
\nonumber\\
&=\frac{4\pi A_{\ell \lambda}}{2\ell+1}\tau_{ij}
\sum_{m_i m_j}\alpha^{m_i}_i \alpha^{m_j}_j
\begin{pmatrix}
1 & 1 & L \\
m_i & m_j & -M
\end{pmatrix}
\frac{(-1)^{\ell-\ell'+M}}{\sqrt{2L+1}}
h_{11L} h_{\ell'\ell L}.
\end{align}

\subsubsection{BipoSH coefficients from $ (\hat{k}\cdot\hat{n})^{\lambda}\tau_{ij}\hat{k}_i\hat{n}_j$  terms}
\begin{align}
(\hat{k}\cdot\hat{n})^{\lambda}\tau_{ij}\hat{k}_i\hat{n}_j
&=\sum_{n=0}^{\lambda} \frac{4\pi A_{n\lambda}}{2n+1}
\sum_{\nu}Y_{n\nu}(\hat{k}) Y^*_{n\nu}(\hat{n})
\left(
\tau_{ij}\sum_{m_i m_j}
\alpha_i^{m_i} Y_{1m_i}(\hat{k})  \alpha_{j}^{m_j} Y_{1m_j}(\hat{n})
\right)
\\
&=\sum_{n=0}^{\lambda} \frac{4\pi A_{n\lambda}}{2n+1}
\tau_{ij}\sum_{\nu}\sum_{m_i m_j}(-1)^{\nu+m_i+m_j}\alpha_i^{m_i} \alpha_j^{m_j} 
\sum_{\ell_3 m_3} \sum_{\ell'_3 m'_3}
Y_{\ell_3 m_3}(\hat{k}) Y_{\ell_3' m_3'}(\hat{n})
\\
&\hspace{3cm}
\times
h_{n1\ell_3} h_{n1\ell'_3}
\begin{pmatrix}
n & 1 & \ell_3 \\
-\nu & -m_i & m_3
\end{pmatrix}
\begin{pmatrix}
n & 1 & \ell'_3 \\
\nu & -m_j & m'_3
\end{pmatrix}
\end{align}
Then, we have
\begin{align}
\pi^{LM}_{\ell\ell'}(k)
&=\int d^2\hat{k}\int d^2\hat{n}~
(\hat{k}\cdot\hat{n})^\lambda \tau_{ij} \hat{k}_i\hat{n}_j
\sum_{mm'}\mathcal{C}^{LM}_{\ell m \ell' m'}Y^*_{\ell m}(\hat{k})Y^*_{\ell' m'}(\hat{n})\nonumber\\
&=\sum_{n=0}^{\lambda} \frac{4\pi A_{n\lambda}}{2n+1}\tau_{ij}
\sum_{\nu}\sum_{m_i m_j}(-1)^{\nu+m_i+m_j}\alpha_i^{m_i} \alpha_j^{m_j} 
h_{n1\ell} h_{n1\ell'}
\sum_{mm'}\mathcal{C}^{LM}_{\ell m \ell' m'}
\begin{pmatrix}
n & 1 & \ell \\
-\nu & -m_i & m
\end{pmatrix}
\begin{pmatrix}
n & 1 & \ell' \\
\nu & -m_j & m'
\end{pmatrix}
\\
&=\sum_{n=0}^{\lambda} \frac{4\pi A_{n\lambda}}{2n+1}(-1)^{n+L+M}
\sqrt{2L+1} \tau_{ij}
\sum_{m_i m_j} \alpha_i^{m_i} \alpha_j^{m_j} 
h_{n1\ell} h_{n1\ell'}
\begin{pmatrix}
1 & 1 & L \\
m_i & m_j & -M
\end{pmatrix}
\begin{Bmatrix}
1 & 1 & L \\
\ell' & \ell & n
\end{Bmatrix}.
\end{align}

\section{\label{app:Identitis}Useful identities}
In this appendix, we summarize the useful properties of the spherical harmonics and the wigner symbols.
\subsection{\label{sec:SH}Spherical harmonics}
The addition theorem of the spherical harmonics tells us that 
\begin{align}
\mathcal{L}_{\ell}(\hat{k}\cdot\hat{x})
=\frac{4\pi}{2\ell+1}\sum_{m} Y_{\ell m}(\hat{k}) Y^*_{\ell m}(\hat{x}),
\end{align}
where $\mathcal{L}_{\ell}(\mu)$ is the $\ell$-th Legendre polynominal.
The orthogonality for the spherical harmonics are 
\begin{align}
&\int d^2\hat{k}~ Y_{\ell m}(\hat{k})Y^*_{\ell' m'}(\hat{k})
=\delta^K_{\ell\ell'}\delta^K_{mm'}\\
&\sum_{\ell m} Y_{\ell m}(\hat{k})Y^*_{\ell m}(\hat{k}')
=\delta^{(2)}_D(\hat{k}-\hat{k}').
\end{align}
The complex conjugate of the spherical harmonics becomes
\begin{align}
Y^*_{\ell m}(\hat{k})
=(-1)^m Y_{\ell -m}(\hat{k}).
\end{align}
A unit vector is written by the spherical harmonics~\cite{Shiraishi:2010kd}
\begin{align}
\hat{k}_i &= \sum_m \alpha^m_i Y_{1m}(\hat{k}), \label{eq:unitvector}\\
{\bm \alpha}^m &= \sqrt{\frac{2\pi}{3}}
\begin{pmatrix}
-m(\delta^K_{m,1}+\delta^K_{m,-1})\\
i(\delta^K_{m,1}+\delta^K_{m,-1})\\
\sqrt{2} \delta^K_{m,0}
\end{pmatrix},
\end{align}
and the coefficient vector ${\bm \alpha}^m$ satisfies the following relations,
\begin{align}
({\bm \alpha}^m )^* &= (-1)^m {\bm \alpha}^{-m} \\
{\bm \alpha}^m  \cdot {\bm \alpha}^{m'} &= \frac{4\pi}{3}(-1)^m\delta^K_{m,-m'},\\
\sum_m \alpha^m_i (\alpha^m_j)^* &= \frac{4\pi}{3}\delta^K_{ij}.
\end{align}
A product of two spherical harmonics which have the same variable is reduced to a spherical harmonics
\begin{align}
Y_{\ell_1 m_1}(\hat{k})Y_{\ell_2 m_2}(\hat{k})
&=\sum_{LM}
\sqrt{\frac{(2\ell_1+1)(2\ell_2+1)(2L+1)}{4\pi}}
\begin{pmatrix}
\ell_1 & \ell_2 & L \\
0   & 0   & 0
\end{pmatrix}
\begin{pmatrix}
\ell_1 & \ell_2 & L \\
m_1 & m_2 & M
\end{pmatrix}
Y^*_{LM}(\hat{k})\nonumber\\
&\equiv\sum_{LM}h_{\ell_1 \ell_2 L}
\begin{pmatrix}
\ell_1 & \ell_2 & L \\
m_1 & m_2 & M
\end{pmatrix}
Y^*_{LM}(\hat{k}).
\label{eq:twoYlm}
\end{align}

\subsection{\label{sec:SH2}Wigner symbols}
The orthogonalities for the Wigner 3-$j$ symbol are
\begin{align}
\sum_{\ell m}
(2\ell+1)
\begin{pmatrix}
\ell_1 & \ell_2 & \ell\\
m_1 & m_2 & m
\end{pmatrix}
\begin{pmatrix}
\ell_1 & \ell_2 & \ell\\
m'_1 & m'_2 & m
\end{pmatrix}
=\delta^K_{m_1 m_1'}\delta^K_{m_2 m_2'},\\
\sum_{m_1 m_2}
\begin{pmatrix}
\ell_1 & \ell_2 & \ell\\
m_1 & m_2 & m
\end{pmatrix}
\begin{pmatrix}
\ell_1 & \ell_2 & \ell'\\
m_1 & m_2 & m'
\end{pmatrix}
=\frac{\delta^K_{\ell \ell'}\delta^K_{m m'}}{2\ell+1}.
\end{align}
The angular momentum coupling implies 
\begin{align}
\sum_m (-1)^{\ell-m}
\begin{pmatrix}
\ell & \ell & L\\
m & -m & M
\end{pmatrix}
=\sqrt{2L+1}\delta_{L,0}\delta_{M,0}.
\end{align}
The wigner 6-$j$ symbol is defined as
\begin{align}
    \begin{Bmatrix}
        \ell_1 & \ell_2 & \ell_3 \\
        \ell_4 & \ell_5 & \ell_6 
    \end{Bmatrix}
    \begin{pmatrix}
        \ell_1 & \ell_2 & \ell_3 \\
        m_1 & m_2 & m_3
    \end{pmatrix}
    \equiv
    \sum_{m_4 m_5 m_6}
    (-1)^{\sum_{i=4}^{6}(\ell_i-m_i)}
    \begin{pmatrix}
        \ell_5  & \ell_1  & \ell_6 \\
        m_5 & -m_1 & -m_6
    \end{pmatrix}
    \begin{pmatrix}
        \ell_6  & \ell_2  & \ell_4 \\
        m_6 & -m_2 & -m_4
    \end{pmatrix}
    \begin{pmatrix}
        \ell_4  & \ell_3  & \ell_5 \\
        m_4 & -m_3 & -m_5
    \end{pmatrix}.
\end{align}

\bibliography{paper}

\begin{thebibliography}{48}%
\makeatletter
\providecommand \@ifxundefined [1]{%
 \@ifx{#1\undefined}
}%
\providecommand \@ifnum [1]{%
 \ifnum #1\expandafter \@firstoftwo
 \else \expandafter \@secondoftwo
 \fi
}%
\providecommand \@ifx [1]{%
 \ifx #1\expandafter \@firstoftwo
 \else \expandafter \@secondoftwo
 \fi
}%
\providecommand \natexlab [1]{#1}%
\providecommand \enquote  [1]{``#1''}%
\providecommand \bibnamefont  [1]{#1}%
\providecommand \bibfnamefont [1]{#1}%
\providecommand \citenamefont [1]{#1}%
\providecommand \href@noop [0]{\@secondoftwo}%
\providecommand \href [0]{\begingroup \@sanitize@url \@href}%
\providecommand \@href[1]{\@@startlink{#1}\@@href}%
\providecommand \@@href[1]{\endgroup#1\@@endlink}%
\providecommand \@sanitize@url [0]{\catcode `\\12\catcode `\$12\catcode
  `\&12\catcode `\#12\catcode `\^12\catcode `\_12\catcode `\%12\relax}%
\providecommand \@@startlink[1]{}%
\providecommand \@@endlink[0]{}%
\providecommand \url  [0]{\begingroup\@sanitize@url \@url }%
\providecommand \@url [1]{\endgroup\@href {#1}{\urlprefix }}%
\providecommand \urlprefix  [0]{URL }%
\providecommand \Eprint [0]{\href }%
\providecommand \doibase [0]{http://dx.doi.org/}%
\providecommand \selectlanguage [0]{\@gobble}%
\providecommand \bibinfo  [0]{\@secondoftwo}%
\providecommand \bibfield  [0]{\@secondoftwo}%
\providecommand \translation [1]{[#1]}%
\providecommand \BibitemOpen [0]{}%
\providecommand \bibitemStop [0]{}%
\providecommand \bibitemNoStop [0]{.\EOS\space}%
\providecommand \EOS [0]{\spacefactor3000\relax}%
\providecommand \BibitemShut  [1]{\csname bibitem#1\endcsname}%
\let\auto@bib@innerbib\@empty
\bibitem [{\citenamefont {Starobinsky}(1980)}]{Starobinsky:1980}%
  \BibitemOpen
  \bibfield  {author} {\bibinfo {author} {\bibfnamefont {A.~A.}\ \bibnamefont
  {Starobinsky}},\ }\href {\doibase 10.1016/0370-2693(80)90670-X} {\bibfield
  {journal} {\bibinfo  {journal} {Phys. Lett.}\ }\textbf {\bibinfo {volume}
  {B91}},\ \bibinfo {pages} {99} (\bibinfo {year} {1980})},\ \bibinfo {note}
  {[,771(1980)]}\BibitemShut {NoStop}%
\bibitem [{\citenamefont {Sato}(1981)}]{Sato:1980}%
  \BibitemOpen
  \bibfield  {author} {\bibinfo {author} {\bibfnamefont {K.}~\bibnamefont
  {Sato}},\ }\href@noop {} {\bibfield  {journal} {\bibinfo  {journal} {Mon.
  Not. Roy. Astron. Soc.}\ }\textbf {\bibinfo {volume} {195}},\ \bibinfo
  {pages} {467} (\bibinfo {year} {1981})}\BibitemShut {NoStop}%
\bibitem [{\citenamefont {Guth}(1981)}]{Guth:1980}%
  \BibitemOpen
  \bibfield  {author} {\bibinfo {author} {\bibfnamefont {A.~H.}\ \bibnamefont
  {Guth}},\ }\href {\doibase 10.1103/PhysRevD.23.347} {\bibfield  {journal}
  {\bibinfo  {journal} {Phys. Rev.}\ }\textbf {\bibinfo {volume} {D23}},\
  \bibinfo {pages} {347} (\bibinfo {year} {1981})},\ \bibinfo {note} {[Adv.
  Ser. Astrophys. Cosmol.3,139(1987)]}\BibitemShut {NoStop}%
\bibitem [{\citenamefont {Alam}\ \emph {et~al.}(2017)\citenamefont {Alam} \emph
  {et~al.}}]{Alam_etal:2016}%
  \BibitemOpen
  \bibfield  {author} {\bibinfo {author} {\bibfnamefont {S.}~\bibnamefont
  {Alam}} \emph {et~al.} (\bibinfo {collaboration} {BOSS}),\ }\href {\doibase
  10.1093/mnras/stx721} {\bibfield  {journal} {\bibinfo  {journal} {Mon. Not.
  Roy. Astron. Soc.}\ }\textbf {\bibinfo {volume} {470}},\ \bibinfo {pages}
  {2617} (\bibinfo {year} {2017})},\ \Eprint {http://arxiv.org/abs/1607.03155}
  {arXiv:1607.03155 [astro-ph.CO]} \BibitemShut {NoStop}%
\bibitem [{\citenamefont {Ellis}\ \emph {et~al.}(2014)\citenamefont {Ellis}
  \emph {et~al.}}]{PFS_wp}%
  \BibitemOpen
  \bibfield  {author} {\bibinfo {author} {\bibfnamefont {R.}~\bibnamefont
  {Ellis}} \emph {et~al.} (\bibinfo {collaboration} {PFS Team}),\ }\href
  {\doibase 10.1093/pasj/pst019} {\bibfield  {journal} {\bibinfo  {journal}
  {Publ. Astron. Soc. Jap.}\ }\textbf {\bibinfo {volume} {66}},\ \bibinfo
  {pages} {R1} (\bibinfo {year} {2014})},\ \Eprint
  {http://arxiv.org/abs/1206.0737} {arXiv:1206.0737 [astro-ph.CO]} \BibitemShut
  {NoStop}%
\bibitem [{\citenamefont {Abell}\ \emph {et~al.}(2009)\citenamefont {Abell}
  \emph {et~al.}}]{LSST_wp}%
  \BibitemOpen
  \bibfield  {author} {\bibinfo {author} {\bibfnamefont {P.~A.}\ \bibnamefont
  {Abell}} \emph {et~al.} (\bibinfo {collaboration} {LSST Science, LSST
  Project}),\ }\href@noop {} {\  (\bibinfo {year} {2009})},\ \Eprint
  {http://arxiv.org/abs/0912.0201} {arXiv:0912.0201 [astro-ph.IM]} \BibitemShut
  {NoStop}%
\bibitem [{\citenamefont {Aghamousa}\ \emph {et~al.}(2016)\citenamefont
  {Aghamousa} \emph {et~al.}}]{DESI_wp}%
  \BibitemOpen
  \bibfield  {author} {\bibinfo {author} {\bibfnamefont {A.}~\bibnamefont
  {Aghamousa}} \emph {et~al.} (\bibinfo {collaboration} {DESI}),\ }\href@noop
  {} {\  (\bibinfo {year} {2016})},\ \Eprint {http://arxiv.org/abs/1611.00036}
  {arXiv:1611.00036 [astro-ph.IM]} \BibitemShut {NoStop}%
\bibitem [{\citenamefont {Dor\'e}\ \emph {et~al.}(2014)\citenamefont {Dor\'e}
  \emph {et~al.}}]{Spherex_wp}%
  \BibitemOpen
  \bibfield  {author} {\bibinfo {author} {\bibfnamefont {O.}~\bibnamefont
  {Dor\'e}} \emph {et~al.},\ }\href@noop {} {\  (\bibinfo {year} {2014})},\
  \Eprint {http://arxiv.org/abs/1412.4872} {arXiv:1412.4872 [astro-ph.CO]}
  \BibitemShut {NoStop}%
\bibitem [{\citenamefont {Bernardeau}\ \emph {et~al.}(2002)\citenamefont
  {Bernardeau}, \citenamefont {Colombi}, \citenamefont {Gaztanaga},\ and\
  \citenamefont {Scoccimarro}}]{PT_review}%
  \BibitemOpen
  \bibfield  {author} {\bibinfo {author} {\bibfnamefont {F.}~\bibnamefont
  {Bernardeau}}, \bibinfo {author} {\bibfnamefont {S.}~\bibnamefont {Colombi}},
  \bibinfo {author} {\bibfnamefont {E.}~\bibnamefont {Gaztanaga}}, \ and\
  \bibinfo {author} {\bibfnamefont {R.}~\bibnamefont {Scoccimarro}},\ }\href
  {\doibase 10.1016/S0370-1573(02)00135-7} {\bibfield  {journal} {\bibinfo
  {journal} {Phys. Rept.}\ }\textbf {\bibinfo {volume} {367}},\ \bibinfo
  {pages} {1} (\bibinfo {year} {2002})},\ \Eprint
  {http://arxiv.org/abs/astro-ph/0112551} {arXiv:astro-ph/0112551 [astro-ph]}
  \BibitemShut {NoStop}%
\bibitem [{\citenamefont {Desjacques}\ \emph {et~al.}(2018)\citenamefont
  {Desjacques}, \citenamefont {Jeong},\ and\ \citenamefont
  {Schmidt}}]{Bias_review}%
  \BibitemOpen
  \bibfield  {author} {\bibinfo {author} {\bibfnamefont {V.}~\bibnamefont
  {Desjacques}}, \bibinfo {author} {\bibfnamefont {D.}~\bibnamefont {Jeong}}, \
  and\ \bibinfo {author} {\bibfnamefont {F.}~\bibnamefont {Schmidt}},\ }\href
  {\doibase 10.1016/j.physrep.2017.12.002} {\bibfield  {journal} {\bibinfo
  {journal} {Phys. Rept.}\ }\textbf {\bibinfo {volume} {733}},\ \bibinfo
  {pages} {1} (\bibinfo {year} {2018})},\ \Eprint
  {http://arxiv.org/abs/1611.09787} {arXiv:1611.09787 [astro-ph.CO]}
  \BibitemShut {NoStop}%
\bibitem [{\citenamefont {Scoccimarro}(2004)}]{Scoccimarro:2004}%
  \BibitemOpen
  \bibfield  {author} {\bibinfo {author} {\bibfnamefont {R.}~\bibnamefont
  {Scoccimarro}},\ }\href {\doibase 10.1103/PhysRevD.70.083007} {\bibfield
  {journal} {\bibinfo  {journal} {Phys. Rev.}\ }\textbf {\bibinfo {volume}
  {D70}},\ \bibinfo {pages} {083007} (\bibinfo {year} {2004})},\ \Eprint
  {http://arxiv.org/abs/astro-ph/0407214} {arXiv:astro-ph/0407214 [astro-ph]}
  \BibitemShut {NoStop}%
\bibitem [{\citenamefont {Takada}\ and\ \citenamefont
  {Hu}(2013)}]{Takada_Hu:2013}%
  \BibitemOpen
  \bibfield  {author} {\bibinfo {author} {\bibfnamefont {M.}~\bibnamefont
  {Takada}}\ and\ \bibinfo {author} {\bibfnamefont {W.}~\bibnamefont {Hu}},\
  }\href {\doibase 10.1103/PhysRevD.87.123504} {\bibfield  {journal} {\bibinfo
  {journal} {Phys. Rev.}\ }\textbf {\bibinfo {volume} {D87}},\ \bibinfo {pages}
  {123504} (\bibinfo {year} {2013})},\ \Eprint {http://arxiv.org/abs/1302.6994}
  {arXiv:1302.6994 [astro-ph.CO]} \BibitemShut {NoStop}%
\bibitem [{\citenamefont {Hamilton}\ \emph {et~al.}(2006)\citenamefont
  {Hamilton}, \citenamefont {Rimes},\ and\ \citenamefont
  {Scoccimarro}}]{Hamilton_etal:2005}%
  \BibitemOpen
  \bibfield  {author} {\bibinfo {author} {\bibfnamefont {A.~J.~S.}\
  \bibnamefont {Hamilton}}, \bibinfo {author} {\bibfnamefont {C.~D.}\
  \bibnamefont {Rimes}}, \ and\ \bibinfo {author} {\bibfnamefont
  {R.}~\bibnamefont {Scoccimarro}},\ }\href {\doibase
  10.1111/j.1365-2966.2006.10709.x} {\bibfield  {journal} {\bibinfo  {journal}
  {Mon. Not. Roy. Astron. Soc.}\ }\textbf {\bibinfo {volume} {371}},\ \bibinfo
  {pages} {1188} (\bibinfo {year} {2006})},\ \Eprint
  {http://arxiv.org/abs/astro-ph/0511416} {arXiv:astro-ph/0511416 [astro-ph]}
  \BibitemShut {NoStop}%
\bibitem [{\citenamefont {Sherwin}\ and\ \citenamefont
  {Zaldarriaga}(2012)}]{Sherwin_Zaldarriaga:2012}%
  \BibitemOpen
  \bibfield  {author} {\bibinfo {author} {\bibfnamefont {B.~D.}\ \bibnamefont
  {Sherwin}}\ and\ \bibinfo {author} {\bibfnamefont {M.}~\bibnamefont
  {Zaldarriaga}},\ }\href {\doibase 10.1103/PhysRevD.85.103523} {\bibfield
  {journal} {\bibinfo  {journal} {Phys. Rev.}\ }\textbf {\bibinfo {volume}
  {D85}},\ \bibinfo {pages} {103523} (\bibinfo {year} {2012})},\ \Eprint
  {http://arxiv.org/abs/1202.3998} {arXiv:1202.3998 [astro-ph.CO]} \BibitemShut
  {NoStop}%
\bibitem [{\citenamefont {Li}\ \emph {et~al.}(2014{\natexlab{a}})\citenamefont
  {Li}, \citenamefont {Hu},\ and\ \citenamefont {Takada}}]{Li_etal:2014a}%
  \BibitemOpen
  \bibfield  {author} {\bibinfo {author} {\bibfnamefont {Y.}~\bibnamefont
  {Li}}, \bibinfo {author} {\bibfnamefont {W.}~\bibnamefont {Hu}}, \ and\
  \bibinfo {author} {\bibfnamefont {M.}~\bibnamefont {Takada}},\ }\href
  {\doibase 10.1103/PhysRevD.89.083519} {\bibfield  {journal} {\bibinfo
  {journal} {Phys. Rev.}\ }\textbf {\bibinfo {volume} {D89}},\ \bibinfo {pages}
  {083519} (\bibinfo {year} {2014}{\natexlab{a}})},\ \Eprint
  {http://arxiv.org/abs/1401.0385} {arXiv:1401.0385 [astro-ph.CO]} \BibitemShut
  {NoStop}%
\bibitem [{\citenamefont {de~Putter}\ \emph {et~al.}(2012)\citenamefont
  {de~Putter}, \citenamefont {Wagner}, \citenamefont {Mena}, \citenamefont
  {Verde},\ and\ \citenamefont {Percival}}]{dePutter_etal:2011}%
  \BibitemOpen
  \bibfield  {author} {\bibinfo {author} {\bibfnamefont {R.}~\bibnamefont
  {de~Putter}}, \bibinfo {author} {\bibfnamefont {C.}~\bibnamefont {Wagner}},
  \bibinfo {author} {\bibfnamefont {O.}~\bibnamefont {Mena}}, \bibinfo {author}
  {\bibfnamefont {L.}~\bibnamefont {Verde}}, \ and\ \bibinfo {author}
  {\bibfnamefont {W.}~\bibnamefont {Percival}},\ }\href {\doibase
  10.1088/1475-7516/2012/04/019} {\bibfield  {journal} {\bibinfo  {journal}
  {JCAP}\ }\textbf {\bibinfo {volume} {1204}},\ \bibinfo {pages} {019}
  (\bibinfo {year} {2012})},\ \Eprint {http://arxiv.org/abs/1111.6596}
  {arXiv:1111.6596 [astro-ph.CO]} \BibitemShut {NoStop}%
\bibitem [{\citenamefont {Li}\ \emph {et~al.}(2014{\natexlab{b}})\citenamefont
  {Li}, \citenamefont {Hu},\ and\ \citenamefont {Takada}}]{Li_etal:2014b}%
  \BibitemOpen
  \bibfield  {author} {\bibinfo {author} {\bibfnamefont {Y.}~\bibnamefont
  {Li}}, \bibinfo {author} {\bibfnamefont {W.}~\bibnamefont {Hu}}, \ and\
  \bibinfo {author} {\bibfnamefont {M.}~\bibnamefont {Takada}},\ }\href
  {\doibase 10.1103/PhysRevD.90.103530} {\bibfield  {journal} {\bibinfo
  {journal} {Phys. Rev.}\ }\textbf {\bibinfo {volume} {D90}},\ \bibinfo {pages}
  {103530} (\bibinfo {year} {2014}{\natexlab{b}})},\ \Eprint
  {http://arxiv.org/abs/1408.1081} {arXiv:1408.1081 [astro-ph.CO]} \BibitemShut
  {NoStop}%
\bibitem [{\citenamefont {Akitsu}\ \emph {et~al.}(2017)\citenamefont {Akitsu},
  \citenamefont {Takada},\ and\ \citenamefont {Li}}]{Akitsu_etal:2016}%
  \BibitemOpen
  \bibfield  {author} {\bibinfo {author} {\bibfnamefont {K.}~\bibnamefont
  {Akitsu}}, \bibinfo {author} {\bibfnamefont {M.}~\bibnamefont {Takada}}, \
  and\ \bibinfo {author} {\bibfnamefont {Y.}~\bibnamefont {Li}},\ }\href
  {\doibase 10.1103/PhysRevD.95.083522} {\bibfield  {journal} {\bibinfo
  {journal} {Phys. Rev.}\ }\textbf {\bibinfo {volume} {D95}},\ \bibinfo {pages}
  {083522} (\bibinfo {year} {2017})},\ \Eprint
  {http://arxiv.org/abs/1611.04723} {arXiv:1611.04723 [astro-ph.CO]}
  \BibitemShut {NoStop}%
\bibitem [{\citenamefont {Barreira}\ \emph {et~al.}(2018)\citenamefont
  {Barreira}, \citenamefont {Krause},\ and\ \citenamefont
  {Schmidt}}]{Barreira_etal:2017}%
  \BibitemOpen
  \bibfield  {author} {\bibinfo {author} {\bibfnamefont {A.}~\bibnamefont
  {Barreira}}, \bibinfo {author} {\bibfnamefont {E.}~\bibnamefont {Krause}}, \
  and\ \bibinfo {author} {\bibfnamefont {F.}~\bibnamefont {Schmidt}},\ }\href
  {\doibase 10.1088/1475-7516/2018/06/015} {\bibfield  {journal} {\bibinfo
  {journal} {JCAP}\ }\textbf {\bibinfo {volume} {1806}},\ \bibinfo {pages}
  {015} (\bibinfo {year} {2018})},\ \Eprint {http://arxiv.org/abs/1711.07467}
  {arXiv:1711.07467 [astro-ph.CO]} \BibitemShut {NoStop}%
\bibitem [{\citenamefont {Schmidt}\ \emph {et~al.}(2018)\citenamefont
  {Schmidt}, \citenamefont {White}, \citenamefont {Schmidt},\ and\
  \citenamefont {Stücker}}]{Schmidt_etal:2018}%
  \BibitemOpen
  \bibfield  {author} {\bibinfo {author} {\bibfnamefont {A.~S.}\ \bibnamefont
  {Schmidt}}, \bibinfo {author} {\bibfnamefont {S.~D.~M.}\ \bibnamefont
  {White}}, \bibinfo {author} {\bibfnamefont {F.}~\bibnamefont {Schmidt}}, \
  and\ \bibinfo {author} {\bibfnamefont {J.}~\bibnamefont {Stücker}},\ }\href
  {\doibase 10.1093/mnras/sty1430} {\bibfield  {journal} {\bibinfo  {journal}
  {Mon. Not. Roy. Astron. Soc.}\ }\textbf {\bibinfo {volume} {479}},\ \bibinfo
  {pages} {162} (\bibinfo {year} {2018})},\ \Eprint
  {http://arxiv.org/abs/1803.03274} {arXiv:1803.03274 [astro-ph.CO]}
  \BibitemShut {NoStop}%
\bibitem [{\citenamefont {Kaiser}(1987)}]{Kaiser:1987}%
  \BibitemOpen
  \bibfield  {author} {\bibinfo {author} {\bibfnamefont {N.}~\bibnamefont
  {Kaiser}},\ }\href@noop {} {\bibfield  {journal} {\bibinfo  {journal} {Mon.
  Not. Roy. Astron. Soc.}\ }\textbf {\bibinfo {volume} {227}},\ \bibinfo
  {pages} {1} (\bibinfo {year} {1987})}\BibitemShut {NoStop}%
\bibitem [{\citenamefont {Alcock}\ and\ \citenamefont
  {Paczynski}(1979)}]{Alcock_Paczynski:1979}%
  \BibitemOpen
  \bibfield  {author} {\bibinfo {author} {\bibfnamefont {C.}~\bibnamefont
  {Alcock}}\ and\ \bibinfo {author} {\bibfnamefont {B.}~\bibnamefont
  {Paczynski}},\ }\href {\doibase 10.1038/281358a0} {\bibfield  {journal}
  {\bibinfo  {journal} {Nature}\ }\textbf {\bibinfo {volume} {281}},\ \bibinfo
  {pages} {358} (\bibinfo {year} {1979})}\BibitemShut {NoStop}%
\bibitem [{\citenamefont {Akitsu}\ and\ \citenamefont
  {Takada}(2018)}]{Akitsu_Takada:2017}%
  \BibitemOpen
  \bibfield  {author} {\bibinfo {author} {\bibfnamefont {K.}~\bibnamefont
  {Akitsu}}\ and\ \bibinfo {author} {\bibfnamefont {M.}~\bibnamefont
  {Takada}},\ }\href {\doibase 10.1103/PhysRevD.97.063527} {\bibfield
  {journal} {\bibinfo  {journal} {Phys. Rev.}\ }\textbf {\bibinfo {volume}
  {D97}},\ \bibinfo {pages} {063527} (\bibinfo {year} {2018})},\ \Eprint
  {http://arxiv.org/abs/1711.00012} {arXiv:1711.00012 [astro-ph.CO]}
  \BibitemShut {NoStop}%
\bibitem [{\citenamefont {Li}\ \emph {et~al.}(2018)\citenamefont {Li},
  \citenamefont {Schmittfull},\ and\ \citenamefont {Seljak}}]{Li_etal:2017}%
  \BibitemOpen
  \bibfield  {author} {\bibinfo {author} {\bibfnamefont {Y.}~\bibnamefont
  {Li}}, \bibinfo {author} {\bibfnamefont {M.}~\bibnamefont {Schmittfull}}, \
  and\ \bibinfo {author} {\bibfnamefont {U.}~\bibnamefont {Seljak}},\ }\href
  {\doibase 10.1088/1475-7516/2018/02/022} {\bibfield  {journal} {\bibinfo
  {journal} {JCAP}\ }\textbf {\bibinfo {volume} {1802}},\ \bibinfo {pages}
  {022} (\bibinfo {year} {2018})},\ \Eprint {http://arxiv.org/abs/1711.00018}
  {arXiv:1711.00018 [astro-ph.CO]} \BibitemShut {NoStop}%
\bibitem [{\citenamefont {Chiang}\ and\ \citenamefont
  {Slosar}(2018)}]{Chiang_Slosar:2018}%
  \BibitemOpen
  \bibfield  {author} {\bibinfo {author} {\bibfnamefont {C.-T.}\ \bibnamefont
  {Chiang}}\ and\ \bibinfo {author} {\bibfnamefont {A.}~\bibnamefont
  {Slosar}},\ }\href@noop {} {\  (\bibinfo {year} {2018})},\ \Eprint
  {http://arxiv.org/abs/1804.02753} {arXiv:1804.02753 [astro-ph.CO]}
  \BibitemShut {NoStop}%
\bibitem [{\citenamefont {Schmidt}\ \emph {et~al.}(2014)\citenamefont
  {Schmidt}, \citenamefont {Pajer},\ and\ \citenamefont
  {Zaldarriaga}}]{Schmidt_etal:2013}%
  \BibitemOpen
  \bibfield  {author} {\bibinfo {author} {\bibfnamefont {F.}~\bibnamefont
  {Schmidt}}, \bibinfo {author} {\bibfnamefont {E.}~\bibnamefont {Pajer}}, \
  and\ \bibinfo {author} {\bibfnamefont {M.}~\bibnamefont {Zaldarriaga}},\
  }\href {\doibase 10.1103/PhysRevD.89.083507} {\bibfield  {journal} {\bibinfo
  {journal} {Phys. Rev.}\ }\textbf {\bibinfo {volume} {D89}},\ \bibinfo {pages}
  {083507} (\bibinfo {year} {2014})},\ \Eprint {http://arxiv.org/abs/1312.5616}
  {arXiv:1312.5616 [astro-ph.CO]} \BibitemShut {NoStop}%
\bibitem [{\citenamefont {Zhu}\ \emph {et~al.}(2016)\citenamefont {Zhu},
  \citenamefont {Pen}, \citenamefont {Yu}, \citenamefont {Er},\ and\
  \citenamefont {Chen}}]{Zhu_etal:2015}%
  \BibitemOpen
  \bibfield  {author} {\bibinfo {author} {\bibfnamefont {H.-M.}\ \bibnamefont
  {Zhu}}, \bibinfo {author} {\bibfnamefont {U.-L.}\ \bibnamefont {Pen}},
  \bibinfo {author} {\bibfnamefont {Y.}~\bibnamefont {Yu}}, \bibinfo {author}
  {\bibfnamefont {X.}~\bibnamefont {Er}}, \ and\ \bibinfo {author}
  {\bibfnamefont {X.}~\bibnamefont {Chen}},\ }\href {\doibase
  10.1103/PhysRevD.93.103504} {\bibfield  {journal} {\bibinfo  {journal} {Phys.
  Rev.}\ }\textbf {\bibinfo {volume} {D93}},\ \bibinfo {pages} {103504}
  (\bibinfo {year} {2016})},\ \Eprint {http://arxiv.org/abs/1511.04680}
  {arXiv:1511.04680 [astro-ph.CO]} \BibitemShut {NoStop}%
\bibitem [{\citenamefont {Barreira}\ and\ \citenamefont
  {Schmidt}(2017)}]{Barreira_Schmidt:2017}%
  \BibitemOpen
  \bibfield  {author} {\bibinfo {author} {\bibfnamefont {A.}~\bibnamefont
  {Barreira}}\ and\ \bibinfo {author} {\bibfnamefont {F.}~\bibnamefont
  {Schmidt}},\ }\href {\doibase 10.1088/1475-7516/2017/06/053} {\bibfield
  {journal} {\bibinfo  {journal} {JCAP}\ }\textbf {\bibinfo {volume} {1706}},\
  \bibinfo {pages} {053} (\bibinfo {year} {2017})},\ \Eprint
  {http://arxiv.org/abs/1703.09212} {arXiv:1703.09212 [astro-ph.CO]}
  \BibitemShut {NoStop}%
\bibitem [{\citenamefont {Varshalovich}\ \emph {et~al.}(1988)\citenamefont
  {Varshalovich}, \citenamefont {Moskalev},\ and\ \citenamefont
  {Khersonsky}}]{book:Varshalovich:1988}%
  \BibitemOpen
  \bibfield  {author} {\bibinfo {author} {\bibfnamefont {D.~A.}\ \bibnamefont
  {Varshalovich}}, \bibinfo {author} {\bibfnamefont {A.~N.}\ \bibnamefont
  {Moskalev}}, \ and\ \bibinfo {author} {\bibfnamefont {V.~K.}\ \bibnamefont
  {Khersonsky}},\ }\href@noop {} {\emph {\bibinfo {title} {{Quantum Theory of
  Angular Momentum: Irreducible Tensors, Spherical Harmonics, Vector Coupling
  Coefficients, 3nj Symbols}}}}\ (\bibinfo  {publisher} {World Scientific},\
  \bibinfo {address} {Singapore},\ \bibinfo {year} {1988})\BibitemShut
  {NoStop}%
\bibitem [{\citenamefont {Shiraishi}\ \emph {et~al.}(2017)\citenamefont
  {Shiraishi}, \citenamefont {Sugiyama},\ and\ \citenamefont
  {Okumura}}]{Shiraishi_etal:2016}%
  \BibitemOpen
  \bibfield  {author} {\bibinfo {author} {\bibfnamefont {M.}~\bibnamefont
  {Shiraishi}}, \bibinfo {author} {\bibfnamefont {N.~S.}\ \bibnamefont
  {Sugiyama}}, \ and\ \bibinfo {author} {\bibfnamefont {T.}~\bibnamefont
  {Okumura}},\ }\href {\doibase 10.1103/PhysRevD.95.063508} {\bibfield
  {journal} {\bibinfo  {journal} {Phys. Rev.}\ }\textbf {\bibinfo {volume}
  {D95}},\ \bibinfo {pages} {063508} (\bibinfo {year} {2017})},\ \Eprint
  {http://arxiv.org/abs/1612.02645} {arXiv:1612.02645 [astro-ph.CO]}
  \BibitemShut {NoStop}%
\bibitem [{\citenamefont {Sugiyama}\ \emph {et~al.}(2018)\citenamefont
  {Sugiyama}, \citenamefont {Shiraishi},\ and\ \citenamefont
  {Okumura}}]{Sugiyama_etal:2017}%
  \BibitemOpen
  \bibfield  {author} {\bibinfo {author} {\bibfnamefont {N.~S.}\ \bibnamefont
  {Sugiyama}}, \bibinfo {author} {\bibfnamefont {M.}~\bibnamefont {Shiraishi}},
  \ and\ \bibinfo {author} {\bibfnamefont {T.}~\bibnamefont {Okumura}},\ }\href
  {\doibase 10.1093/mnras/stx2333} {\bibfield  {journal} {\bibinfo  {journal}
  {Mon. Not. Roy. Astron. Soc.}\ }\textbf {\bibinfo {volume} {473}},\ \bibinfo
  {pages} {2737} (\bibinfo {year} {2018})},\ \Eprint
  {http://arxiv.org/abs/1704.02868} {arXiv:1704.02868 [astro-ph.CO]}
  \BibitemShut {NoStop}%
\bibitem [{\citenamefont {Bartolo}\ \emph {et~al.}(2018)\citenamefont
  {Bartolo}, \citenamefont {Kehagias}, \citenamefont {Liguori}, \citenamefont
  {Riotto}, \citenamefont {Shiraishi},\ and\ \citenamefont
  {Tansella}}]{Bartolo:2017sbu}%
  \BibitemOpen
  \bibfield  {author} {\bibinfo {author} {\bibfnamefont {N.}~\bibnamefont
  {Bartolo}}, \bibinfo {author} {\bibfnamefont {A.}~\bibnamefont {Kehagias}},
  \bibinfo {author} {\bibfnamefont {M.}~\bibnamefont {Liguori}}, \bibinfo
  {author} {\bibfnamefont {A.}~\bibnamefont {Riotto}}, \bibinfo {author}
  {\bibfnamefont {M.}~\bibnamefont {Shiraishi}}, \ and\ \bibinfo {author}
  {\bibfnamefont {V.}~\bibnamefont {Tansella}},\ }\href {\doibase
  10.1103/PhysRevD.97.023503} {\bibfield  {journal} {\bibinfo  {journal} {Phys.
  Rev.}\ }\textbf {\bibinfo {volume} {D97}},\ \bibinfo {pages} {023503}
  (\bibinfo {year} {2018})},\ \Eprint {http://arxiv.org/abs/1709.05695}
  {arXiv:1709.05695 [astro-ph.CO]} \BibitemShut {NoStop}%
\bibitem [{\citenamefont {Szalay}\ \emph {et~al.}(1998)\citenamefont {Szalay},
  \citenamefont {Matsubara},\ and\ \citenamefont {Landy}}]{Szalay:1997cc}%
  \BibitemOpen
  \bibfield  {author} {\bibinfo {author} {\bibfnamefont {A.~S.}\ \bibnamefont
  {Szalay}}, \bibinfo {author} {\bibfnamefont {T.}~\bibnamefont {Matsubara}}, \
  and\ \bibinfo {author} {\bibfnamefont {S.~D.}\ \bibnamefont {Landy}},\ }\href
  {\doibase 10.1086/311293} {\bibfield  {journal} {\bibinfo  {journal}
  {Astrophys. J.}\ }\textbf {\bibinfo {volume} {498}},\ \bibinfo {pages} {L1}
  (\bibinfo {year} {1998})},\ \Eprint {http://arxiv.org/abs/astro-ph/9712007}
  {arXiv:astro-ph/9712007 [astro-ph]} \BibitemShut {NoStop}%
\bibitem [{\citenamefont {Szapudi}(2004)}]{Szapudi:2004gh}%
  \BibitemOpen
  \bibfield  {author} {\bibinfo {author} {\bibfnamefont {I.}~\bibnamefont
  {Szapudi}},\ }\href {\doibase 10.1086/423168} {\bibfield  {journal} {\bibinfo
   {journal} {Astrophys. J.}\ }\textbf {\bibinfo {volume} {614}},\ \bibinfo
  {pages} {51} (\bibinfo {year} {2004})},\ \Eprint
  {http://arxiv.org/abs/astro-ph/0404477} {arXiv:astro-ph/0404477 [astro-ph]}
  \BibitemShut {NoStop}%
\bibitem [{\citenamefont {Papai}\ and\ \citenamefont
  {Szapudi}(2008)}]{Papai:2008bd}%
  \BibitemOpen
  \bibfield  {author} {\bibinfo {author} {\bibfnamefont {P.}~\bibnamefont
  {Papai}}\ and\ \bibinfo {author} {\bibfnamefont {I.}~\bibnamefont
  {Szapudi}},\ }\href {\doibase 10.1111/j.1365-2966.2008.13572.x} {\bibfield
  {journal} {\bibinfo  {journal} {Mon. Not. Roy. Astron. Soc.}\ }\textbf
  {\bibinfo {volume} {389}},\ \bibinfo {pages} {292} (\bibinfo {year}
  {2008})},\ \Eprint {http://arxiv.org/abs/0802.2940} {arXiv:0802.2940
  [astro-ph]} \BibitemShut {NoStop}%
\bibitem [{\citenamefont {Bertacca}\ \emph {et~al.}(2012)\citenamefont
  {Bertacca}, \citenamefont {Maartens}, \citenamefont {Raccanelli},\ and\
  \citenamefont {Clarkson}}]{Bertacca:2012tp}%
  \BibitemOpen
  \bibfield  {author} {\bibinfo {author} {\bibfnamefont {D.}~\bibnamefont
  {Bertacca}}, \bibinfo {author} {\bibfnamefont {R.}~\bibnamefont {Maartens}},
  \bibinfo {author} {\bibfnamefont {A.}~\bibnamefont {Raccanelli}}, \ and\
  \bibinfo {author} {\bibfnamefont {C.}~\bibnamefont {Clarkson}},\ }\href
  {\doibase 10.1088/1475-7516/2012/10/025} {\bibfield  {journal} {\bibinfo
  {journal} {JCAP}\ }\textbf {\bibinfo {volume} {1210}},\ \bibinfo {pages}
  {025} (\bibinfo {year} {2012})},\ \Eprint {http://arxiv.org/abs/1205.5221}
  {arXiv:1205.5221 [astro-ph.CO]} \BibitemShut {NoStop}%
\bibitem [{\citenamefont {Raccanelli}\ \emph {et~al.}(2014)\citenamefont
  {Raccanelli}, \citenamefont {Bertacca}, \citenamefont {Doré},\ and\
  \citenamefont {Maartens}}]{Raccanelli:2013dza}%
  \BibitemOpen
  \bibfield  {author} {\bibinfo {author} {\bibfnamefont {A.}~\bibnamefont
  {Raccanelli}}, \bibinfo {author} {\bibfnamefont {D.}~\bibnamefont
  {Bertacca}}, \bibinfo {author} {\bibfnamefont {O.}~\bibnamefont {Doré}}, \
  and\ \bibinfo {author} {\bibfnamefont {R.}~\bibnamefont {Maartens}},\ }\href
  {\doibase 10.1088/1475-7516/2014/08/022} {\bibfield  {journal} {\bibinfo
  {journal} {JCAP}\ }\textbf {\bibinfo {volume} {1408}},\ \bibinfo {pages}
  {022} (\bibinfo {year} {2014})},\ \Eprint {http://arxiv.org/abs/1306.6646}
  {arXiv:1306.6646 [astro-ph.CO]} \BibitemShut {NoStop}%
\bibitem [{\citenamefont {Sugiyama}\ \emph {et~al.}(2019)\citenamefont
  {Sugiyama}, \citenamefont {Saito}, \citenamefont {Beutler},\ and\
  \citenamefont {Seo}}]{Sugiyama:2018yzo}%
  \BibitemOpen
  \bibfield  {author} {\bibinfo {author} {\bibfnamefont {N.~S.}\ \bibnamefont
  {Sugiyama}}, \bibinfo {author} {\bibfnamefont {S.}~\bibnamefont {Saito}},
  \bibinfo {author} {\bibfnamefont {F.}~\bibnamefont {Beutler}}, \ and\
  \bibinfo {author} {\bibfnamefont {H.-J.}\ \bibnamefont {Seo}},\ }\href
  {\doibase 10.1093/mnras/sty3249} {\bibfield  {journal} {\bibinfo  {journal}
  {Mon. Not. Roy. Astron. Soc.}\ }\textbf {\bibinfo {volume} {484}},\ \bibinfo
  {pages} {364} (\bibinfo {year} {2019})},\ \Eprint
  {http://arxiv.org/abs/1803.02132} {arXiv:1803.02132 [astro-ph.CO]}
  \BibitemShut {NoStop}%
\bibitem [{\citenamefont {Dai}\ \emph {et~al.}(2015)\citenamefont {Dai},
  \citenamefont {Pajer},\ and\ \citenamefont {Schmidt}}]{Dai_etal:2015}%
  \BibitemOpen
  \bibfield  {author} {\bibinfo {author} {\bibfnamefont {L.}~\bibnamefont
  {Dai}}, \bibinfo {author} {\bibfnamefont {E.}~\bibnamefont {Pajer}}, \ and\
  \bibinfo {author} {\bibfnamefont {F.}~\bibnamefont {Schmidt}},\ }\href
  {\doibase 10.1088/1475-7516/2015/10/059} {\bibfield  {journal} {\bibinfo
  {journal} {JCAP}\ }\textbf {\bibinfo {volume} {1510}},\ \bibinfo {pages}
  {059} (\bibinfo {year} {2015})},\ \Eprint {http://arxiv.org/abs/1504.00351}
  {arXiv:1504.00351 [astro-ph.CO]} \BibitemShut {NoStop}%
\bibitem [{\citenamefont {Ip}\ and\ \citenamefont
  {Schmidt}(2017)}]{Ip_Schmidt:2016}%
  \BibitemOpen
  \bibfield  {author} {\bibinfo {author} {\bibfnamefont {H.~Y.}\ \bibnamefont
  {Ip}}\ and\ \bibinfo {author} {\bibfnamefont {F.}~\bibnamefont {Schmidt}},\
  }\href {\doibase 10.1088/1475-7516/2017/02/025} {\bibfield  {journal}
  {\bibinfo  {journal} {JCAP}\ }\textbf {\bibinfo {volume} {1702}},\ \bibinfo
  {pages} {025} (\bibinfo {year} {2017})},\ \Eprint
  {http://arxiv.org/abs/1610.01059} {arXiv:1610.01059 [astro-ph.CO]}
  \BibitemShut {NoStop}%
\bibitem [{\citenamefont {Aghanim}\ \emph {et~al.}(2018)\citenamefont {Aghanim}
  \emph {et~al.}}]{Planck2018_cosmo}%
  \BibitemOpen
  \bibfield  {author} {\bibinfo {author} {\bibfnamefont {N.}~\bibnamefont
  {Aghanim}} \emph {et~al.} (\bibinfo {collaboration} {Planck}),\ }\href@noop
  {} {\  (\bibinfo {year} {2018})},\ \Eprint {http://arxiv.org/abs/1807.06209}
  {arXiv:1807.06209 [astro-ph.CO]} \BibitemShut {NoStop}%
\bibitem [{\citenamefont {Padmanabhan}\ and\ \citenamefont
  {White}(2008)}]{Padmanabhan_etal:2008}%
  \BibitemOpen
  \bibfield  {author} {\bibinfo {author} {\bibfnamefont {N.}~\bibnamefont
  {Padmanabhan}}\ and\ \bibinfo {author} {\bibfnamefont {M.~J.}\ \bibnamefont
  {White}},\ }\href {\doibase 10.1103/PhysRevD.77.123540} {\bibfield  {journal}
  {\bibinfo  {journal} {Phys. Rev.}\ }\textbf {\bibinfo {volume} {D77}},\
  \bibinfo {pages} {123540} (\bibinfo {year} {2008})},\ \Eprint
  {http://arxiv.org/abs/0804.0799} {arXiv:0804.0799 [astro-ph]} \BibitemShut
  {NoStop}%
\bibitem [{\citenamefont {Sirko}(2005)}]{Sirko:2005}%
  \BibitemOpen
  \bibfield  {author} {\bibinfo {author} {\bibfnamefont {E.}~\bibnamefont
  {Sirko}},\ }\href {\doibase 10.1086/497090} {\bibfield  {journal} {\bibinfo
  {journal} {Astrophys. J.}\ }\textbf {\bibinfo {volume} {634}},\ \bibinfo
  {pages} {728} (\bibinfo {year} {2005})},\ \Eprint
  {http://arxiv.org/abs/astro-ph/0503106} {arXiv:astro-ph/0503106 [astro-ph]}
  \BibitemShut {NoStop}%
\bibitem [{\citenamefont {Wagner}\ \emph {et~al.}(2015)\citenamefont {Wagner},
  \citenamefont {Schmidt}, \citenamefont {Chiang},\ and\ \citenamefont
  {Komatsu}}]{Wagner_etal:2014}%
  \BibitemOpen
  \bibfield  {author} {\bibinfo {author} {\bibfnamefont {C.}~\bibnamefont
  {Wagner}}, \bibinfo {author} {\bibfnamefont {F.}~\bibnamefont {Schmidt}},
  \bibinfo {author} {\bibfnamefont {C.-T.}\ \bibnamefont {Chiang}}, \ and\
  \bibinfo {author} {\bibfnamefont {E.}~\bibnamefont {Komatsu}},\ }\href
  {\doibase 10.1093/mnrasl/slu187} {\bibfield  {journal} {\bibinfo  {journal}
  {Mon. Not. Roy. Astron. Soc.}\ }\textbf {\bibinfo {volume} {448}},\ \bibinfo
  {pages} {L11} (\bibinfo {year} {2015})},\ \Eprint
  {http://arxiv.org/abs/1409.6294} {arXiv:1409.6294 [astro-ph.CO]} \BibitemShut
  {NoStop}%
\bibitem [{\citenamefont {Baldauf}\ \emph {et~al.}(2016)\citenamefont
  {Baldauf}, \citenamefont {Seljak}, \citenamefont {Senatore},\ and\
  \citenamefont {Zaldarriaga}}]{Baldauf_etal:2015}%
  \BibitemOpen
  \bibfield  {author} {\bibinfo {author} {\bibfnamefont {T.}~\bibnamefont
  {Baldauf}}, \bibinfo {author} {\bibfnamefont {U.}~\bibnamefont {Seljak}},
  \bibinfo {author} {\bibfnamefont {L.}~\bibnamefont {Senatore}}, \ and\
  \bibinfo {author} {\bibfnamefont {M.}~\bibnamefont {Zaldarriaga}},\ }\href
  {\doibase 10.1088/1475-7516/2016/09/007} {\bibfield  {journal} {\bibinfo
  {journal} {JCAP}\ }\textbf {\bibinfo {volume} {1609}},\ \bibinfo {pages}
  {007} (\bibinfo {year} {2016})},\ \Eprint {http://arxiv.org/abs/1511.01465}
  {arXiv:1511.01465 [astro-ph.CO]} \BibitemShut {NoStop}%
\bibitem [{\citenamefont {Jamieson}\ and\ \citenamefont
  {LoVerde}(2018)}]{Jamieson_LoVerde:2018}%
  \BibitemOpen
  \bibfield  {author} {\bibinfo {author} {\bibfnamefont {D.}~\bibnamefont
  {Jamieson}}\ and\ \bibinfo {author} {\bibfnamefont {M.}~\bibnamefont
  {LoVerde}},\ }\href@noop {} {\  (\bibinfo {year} {2018})},\ \Eprint
  {http://arxiv.org/abs/1812.08765} {arXiv:1812.08765 [astro-ph.CO]}
  \BibitemShut {NoStop}%
\bibitem [{\citenamefont {Nan}\ \emph {et~al.}(2019)\citenamefont {Nan},
  \citenamefont {Yamamoto}, \citenamefont {Aoki}, \citenamefont {Iso},\ and\
  \citenamefont {Yamauchi}}]{Nan_etal:2019}%
  \BibitemOpen
  \bibfield  {author} {\bibinfo {author} {\bibfnamefont {Y.}~\bibnamefont
  {Nan}}, \bibinfo {author} {\bibfnamefont {K.}~\bibnamefont {Yamamoto}},
  \bibinfo {author} {\bibfnamefont {H.}~\bibnamefont {Aoki}}, \bibinfo {author}
  {\bibfnamefont {S.}~\bibnamefont {Iso}}, \ and\ \bibinfo {author}
  {\bibfnamefont {D.}~\bibnamefont {Yamauchi}},\ }\href {\doibase
  10.1103/PhysRevD.99.103512} {\bibfield  {journal} {\bibinfo  {journal} {Phys.
  Rev.}\ }\textbf {\bibinfo {volume} {D99}},\ \bibinfo {pages} {103512}
  (\bibinfo {year} {2019})},\ \Eprint {http://arxiv.org/abs/1901.11181}
  {arXiv:1901.11181 [astro-ph.CO]} \BibitemShut {NoStop}%
\bibitem [{\citenamefont {Shiraishi}\ \emph {et~al.}(2011)\citenamefont
  {Shiraishi}, \citenamefont {Nitta}, \citenamefont {Yokoyama}, \citenamefont
  {Ichiki},\ and\ \citenamefont {Takahashi}}]{Shiraishi:2010kd}%
  \BibitemOpen
  \bibfield  {author} {\bibinfo {author} {\bibfnamefont {M.}~\bibnamefont
  {Shiraishi}}, \bibinfo {author} {\bibfnamefont {D.}~\bibnamefont {Nitta}},
  \bibinfo {author} {\bibfnamefont {S.}~\bibnamefont {Yokoyama}}, \bibinfo
  {author} {\bibfnamefont {K.}~\bibnamefont {Ichiki}}, \ and\ \bibinfo {author}
  {\bibfnamefont {K.}~\bibnamefont {Takahashi}},\ }\href {\doibase
  10.1143/PTP.125.795} {\bibfield  {journal} {\bibinfo  {journal} {Prog. Theor.
  Phys.}\ }\textbf {\bibinfo {volume} {125}},\ \bibinfo {pages} {795} (\bibinfo
  {year} {2011})},\ \Eprint {http://arxiv.org/abs/1012.1079} {arXiv:1012.1079
  [astro-ph.CO]} \BibitemShut {NoStop}%
\end{thebibliography}%

\end{document}